\documentstyle[a4,12pt,epsf,style]{book}

\def \L {{\cal L}}

\def \K {{\cal K}}
\def \del {\partial}
\def \4pmeasure#1{\frac{d^4 #1}{(2\pi)^4}}
\def \3pmeasure#1{\frac{d^3 \mbox{\bf #1}}{(2\pi)^3}}

\def \Vec#1{\mbox{{\boldmath $#1$}}}

\def \psib{\bar{\psi}}

\def \g5{\gamma_5}
\def \dsla {\partial \!\!\! /}

\def \qsla {q \!\!\! /}
\def \ksla {k \!\!\! /}
\def \Psla {P \!\!\!\! /}
\def \Dsla {D \!\!\!\! /}

\def \Tr {\mbox{Tr}}
\def \tr {\mbox{tr}}

\def \flavor {\frac{\lambda^{\alpha}}{2}}

\def \GeV {\mbox{GeV}}
\def \MeV {\mbox{MeV}}

\def \lamQ {\Lambda_{\rm QCD}}
\def \lamUV {\Lambda_{\rm UV}}
\def \tIF {t_{\rm IF}}

\newcommand{\UA}{\mbox{$U_{A}(1)$}}
\newcommand{\qbar}{\mbox{$\bar q$}}
\newcommand{\ubar}{\mbox{$\bar u$}}
\newcommand{\dbar}{\mbox{$\bar d$}}
\newcommand{\sbar}{\mbox{$\bar s$}}
\newcommand{\ie}{{\it i.e.}}
\newcommand{\bea}{\begin{eqnarray}}
\newcommand{\eea}{\end{eqnarray}}
\newcommand{\la}{\left\langle}
\newcommand{\ra}{\right\rangle}

\def\JLone<#1,#2>{#1}
\def\JLtwo<#1,#2,#3>{#2}
\def\JLyear<#1,#2,#3,#4>{#3}
\def\JLpage<#1,#2,#3,#4>{#4}
\newcommand\JL[1]{\JLone<#1>\ {\bf \JLtwo<#1>} (\JLyear<#1>), \JLpage<#1>}
\def\Jpage<#1,#2,#3>{#3}

\newcommand\andvol[1]{{\bf \JLone<#1>} (\JLtwo<#1>), \Jpage<#1>}

\newcommand\PTP[1]{Prog.\ Theor.\ Phys.\ \andvol{#1}}

\newcommand\PRC[1]{Phys.\ Rev.\ C\ \andvol{#1}}
\newcommand\PRD[1]{Phys.\ Rev.\ D\ \andvol{#1}}

\newcommand\PRL[1]{Phys.\ Rev.\ Lett.\ \andvol{#1}}

\newcommand\PLB[1]{Phys.\ Lett.\ B\ \andvol{#1}}
\newcommand\NPA[1]{Nucl.\ Phys.\ A\ \andvol{#1}}
\newcommand\NPB[1]{Nucl.\ Phys.\ B\ \andvol{#1}}

\newcommand\PRP[1]{Phys. Rep.\ \andvol{#1}}

\def\citen#1{%
\if@filesw \immediate \write \@auxout {\string \citation {#1}}\fi 
\@tempcntb\m@ne \let\@h@ld\relax \def\@citea{}%
\@for \@citeb:=#1\do {%
  \@ifundefined {b@\@citeb}%
    {\@h@ld\@citea\@tempcntb\m@ne{\bf ?}%
    \@warning {Citation `\@citeb ' on page \thepage \space undefined}}%
    {\@tempcnta\@tempcntb \advance\@tempcnta\@ne
    \setbox\z@\hbox\bgroup 
    \ifnum0<0\csname b@\@citeb \endcsname) \relax
       \egroup \@tempcntb\number\csname b@\@citeb \endcsname \relax
       \else \egroup \@tempcntb\m@ne \fi
    \ifnum\@tempcnta=\@tempcntb 
       \ifx\@h@ld\relax 
          \edef \@h@ld{\@citea\csname b@\@citeb\endcsname)}%
       \else 
          \edef\@h@ld{\penalty\@highpenalty\hskip.15em plus.1em minus.1em
            \hbox{-}\penalty\@highpenalty\hskip.15em plus.1em minus.1em
            \csname b@\@citeb\endcsname)}%
       \fi
    \else   
       \@h@ld\@citea\csname b@\@citeb \endcsname)\let\@h@ld\relax
    \fi}%
 \def\@citea{,\penalty\@highpenalty\hskip.15em plus.1em minus.1em}%
}\@h@ld}
\def\@citex[#1]#2{\@cite{\citen{#2}}{#1}}%
\def\@cite#1#2{\leavevmode\unskip
  \ifnum\lastpenalty=\z@\penalty\@highpenalty\fi
   $^{\hskip.15em plus.1em minus.1em \multiply\@highpenalty 3 #1%
      \if@tempswa,\penalty\@highpenalty\ #2\fi 
    }$\spacefactor\@m}
\begin{document}


\setlength{\baselineskip}{18pt}

\begin{titlepage}

\begin{center}

{\huge

 Light Scalar Mesons

in the Improved Ladder Approximation 

of QCD 

with Strong $U_A(1)$ Breaking
 }
\vspace{6cm} \\
\Large

Toru Umekawa

\vspace{4cm}
Thesis submitted for the Degree 

of Doctor of Science 

\vspace{1cm}

Department of Physics, Tokyo Institute of Technology, 

Meguro, Tokyo, 152-8551 Japan 

\vspace{1cm}

January, 2003

\end{center}
\end{titlepage}
\newpage
\thispagestyle{empty}

{\Large \bf Abstract}

\vspace{1cm}

The spectrum and the mixing angle of the light scalar nonet 
mesons are studied using the extended Nambu-Jona-Lasinio (NJL) 
model as well as the improved ladder approximation (ILA) of 
QCD with \UA\  symmetry breaking interaction. 
The ILA is an approximation that is consistent with chiral 
symmetry and consists of the rainbow approximation of the 
Schwinger-Dyson equation and the ladder approximation 
of the Bethe-Salpeter equation. Improvement is 
made by the use of the running coupling constant 
according to the Higashijima-Miransky method. 
The \UA\  breaking is supposed to come from the coupling 
of light quarks to instanton, and is represented by a 
contact six-quark vertex, called the 
Kobayashi-Maskawa-'t Hooft (KMT) interaction. 
The strength of the KMT interaction in the NJL model is 
determined so as to reproduce the electromagnetic decays 
of the $\eta$ meson. 
That in the ILA approach is 
determined so as to reproduce the pseudoscalar meson 
spectrum. 
This interaction is so strong that it causes large mixing 
of $(u\bar{u}+d\bar{d})$ and $s\bar{s}$, in the $\eta$ meson. 

In the extended NJL model, we study the 
qualitative features of the 
scalar meson spectrum. 
In the scalar nonet spectrum, the KMT interaction is 
found to give the right ordering of $\sigma - a_0$ 
masses and a few hundred MeV mass difference 
between the $\sigma$ and $a_0$ mesons. 
We also find that the strangeness content in the 
$\sigma$ meson is about $15\%$. 

In the ILA approach, we confirm the qualitative features 
of the results from the extended NJL. 
We obtain the mass spectra of the light 
pseudoscalar nonet mesons and 
the $\sigma, a_0$ mesons which are consistent with current 
experimental data. 
They are both reproduced with the same parameter choice. 
On the other hand, we show that 
the obtained $f_0$ and $K_0^*$ state can not be 
identified directly with the experimental data. 
It may suggest that the $f_0(980)$ and $\kappa(700-900)$ 
states may not be explained only by the 
quark-anti quark state. 

We also find that the strangeness content in the 
$\sigma$ meson is about $5\%$. 
We obtain the result that the $U_A(1)$ breaking interaction 
reproducing the mass spectrum of the pseudoscalar meson 
gives large effects on the scalar meson mass spectrum. 

\newpage

\pagenumbering{roman}
\setcounter{page}{1}
 
\tableofcontents
\listoffigures
\listoftables
\newpage
\pagenumbering{arabic}
\setcounter{page}{0}

\chapter{Introduction}

Understanding low-lying hadron spectrum is one of the most challenging 
problems in the quantum chromodynamics (QCD).  
The spectrum is highly nontrivial due to the nonperturbative 
complexity, such as dynamical chiral symmetry breaking (D$\chi$SB),
axial U(1) anomaly.  
For instance, the pseudoscalar mesons are off-scale light, if we 
suppose that they are bound states of a quark and an antiquark, 
while their spin-flip partners, \ie, vector mesons, are normal with
masses about 2/3 of the baryon masses.
This ``anomaly'' in the pseudoscalar mesons is attributed to 
their Nambu-Goldstone-boson nature associated with the 
D$\chi$SB.
This is strongly in contrast with the heavy meson spectroscopy,
such as heavy quarkonia and heavy flavor mesons.  There the 
spectrum is more like the hydrogen atom with slightly
stronger fine and hyperfine splittings.

It should be noticed that the low-lying hadrons are the key to
explore the complicated QCD vacuum, as in QCD we are not able to
``measure'' the bulk properties of the ground state, which 
can be accessed directly in the case of condensed matter physics.
Thus it is important to explore the properties of the low-lying
hadrons from the viewpoints of QCD dynamics and symmetries.

Another nontrivial effect comes from \UA\ symmetry, which is
expected to be broken by anomaly.
Weinberg showed that the mass of $\eta'$ should be less
than $\sqrt{3} m_{\pi}$ if \UA\ symmetry were not
explicitly broken.\cite{Weinberg1975}
Thus the \UA\ symmetry must be broken.
Later, 't Hooft pointed out the relation between 
\UA\ anomaly and topological gluon configurations of QCD and 
showed that the interaction of light quarks and instantons breaks 
the \UA\ symmetry.\cite{tHooft1976}
He also showed that such interaction can be represented by a
local $2N_{f}$ quark vertex, which is antisymmetric under
flavor exchanges, in the dilute instanton gas approximation.
The dynamics of instantons in the multi-instanton vacuum has
been studied by many authors, either in the models or in the
lattice QCD approach, and the widely accepted picture is that
the QCD vacuum consists of small instantons of the size about
1/3 fm with the density of 1 instanton (or anti-instanton) per
fm$^{4}$.\cite{Instanton}

According to such instanton vacuum picture, the hadron spectrum 
shows its signature.  The $\eta-\eta'$ mass difference is the obvious one, 
which can be understood by flavor mixing in the $I=0$ 
${1\over \sqrt{2}} \left( u\ubar+d\dbar \right)$ 
and $s\sbar$. 
Without the flavor mixing, ${1\over \sqrt{2}} \left( u\ubar+d\dbar \right)$ 
and $s\sbar$ would form 
mass eigenstates, and thus the ideal mixing is achieved. 
This is natural if the Okubo-Zweig-Iizuka (OZI) rule applies.
However, the OZI rule is known to be significantly broken in 
the pseudoscalar mesons.
For instance in the 
Nambu-Jona-Lasinio(NJL) model,  the electromagnetic $\eta$ decay 
processes suggest  that the mixing of 
${1\over \sqrt{2}} \left( u\ubar+d\dbar \right)$ 
and $s\sbar$ is indeed strong so that the $\eta$ meson is close to  
the pure octet state.\cite{TNO1997}

Recently, the scalar mesons, $J^{\pi}=0^{+}$, attracts a lot of 
attention by two reasons.\cite{sigmaworkshop}
(1) Experimental evidence for $\sigma$ 
($I=0$) scalar meson of mass around 500-800 MeV is overwhelming.\cite{PDG2002,BES}
Especially the decays of heavy mesons show clear peaks in the 
$\pi\pi$ invariant mass spectrum.
(2) The roles of the scalar mesons in chiral symmetry have been 
stressed in the context of high temperature and/or density hadronic 
matter.\cite{HK1985}
It is believed that chiral symmetry will be restored in the QCD ground 
state at high temperature (and/or baryon density).  Above the critical 
temperature of order 150 MeV the world is nearly chiral symmetric and 
we expect that hadrons belong to an irreducible representation of 
chiral symmetry, if we neglect small mixing due to finite quark mass.
The pion is not any more a Nambu-Goldstone(NG) boson, and has a finite 
mass and should be degenerate with a scalar meson, \ie, $\sigma$. 

For $N_f=3$, we expect to have a $U(3)$ nonet of light scalar mesons, 
which are the chiral partners of the pseudoscalar nonet. 
We assume that $\sigma$ is the member of this nonet. 
The naive assignment of the flavor component of $\sigma(I=0)$ and 
$a_0(I=1)$ 
is $u\bar{u} + d\bar{d}$ and $u\bar{u} - d\bar{d}$, respectively. 
The masses of the $u\bar{u} + d\bar{d}$ and $u\bar{u} - d\bar{d}$
mesons  must become degenerate, if the flavor mixing interaction 
is absent. 
However, the mass of $a_0(980)$, which is the lightest $a_0$ meson, 
is larger than that of $\sigma$ by about $400 [\MeV]$. 
We consider that this mass splitting is explained again by the $U_A(1)$ 
breaking interaction. 
As is in the case of the light pseudoscalar mesons, 
the $U_A(1)$ breaking interaction contributes significantly  
for the light scalar meson nonet and then $\sigma - a_0$ mass 
splitting is explained by the flavor mixing. 
The goal  of this paper is to study  the 
role of the \UA\  breaking interaction 
to the mass spectrum  of the light nonet scalar meson. 
Here we study the lowest scalar meson states in the $I = 0, \, 1, \, 1/2$ 
channels and the second lowest one in $I = 0$ channel which we call 
 $\sigma$, $a_0$, $K_0^*$ and $f_0$, respectively. 
The identification of these states with the 
experimentally observed states will be given later.

We first employ a simple model, 
the extended Nambu-Jona-Lasinio(NJL) model. 
It is the simplest possible quark model 
with the correct symmetry 
structure for the present purposes. \cite{NJL}  The chiral symmetry is broken 
both explicitly by the quark mass term and dynamically by quark loops,
while the \UA\ symmetry is broken by the Kobayashi-Maskawa-'t Hooft (KMT) 
interaction. 
\cite{tHooft1976,KM1970}
The KMT interaction is a contact six-quark interaction, 
which 
represents the \UA\  breaking effects 
induced by the instanton configuration of the gluon in the 
vacuum. 

After understanding the qualitative features of the 
scalar meson spectrum in the extended NJL model, 
we proceed to the analysis using 
the improved ladder approximation (ILA) \cite{ILA}
of QCD with the KMT interaction. 
The ILA of QCD is 
the approximation in which the gluon is integrated out 
and is written only in terms of the quark degrees of freedom. 
The ILA is an approximation that is consistent with chiral 
symmetry. 
We employ the rainbow approximation of the 
Schwinger-Dyson (SD) equation for the quark propagator 
and the ladder approximation 
of the Bethe-Salpeter (BS) equation for the quark-anti quark 
bound state. 
It is important to note that these two approximated 
equations are consistent with the chiral symmetry. 
Improvement is 
made by the use of the running coupling constant 
according to the Higashijima-Miransky \cite{Higashijima} method. 
The KMT interaction is introduced in the ILA 
as an explicit interaction kernel. 
Naito et al. \cite{nairon3,NNTYO2000}  showed explicitly that 
the chiral symmetry is indeed preserved in this 
formulation and properties of the pseudoscalar 
meson, such as the masses and decay constants, are 
consistent with those of the Nambu-Goldstone (NG)
bosons of chiral symmetry breaking. 
They furthermore considered effects of the $U_A(1)$ 
breaking and explicit chiral symmetry breaking due to 
the finite quark masses. 
They successfully reproduced the realistic 
mass spectrum of the pseudoscalar mesons. 

In this thesis, we consider the scalar mesons in the same 
formulation. 
The advantage of this approach is consistency with 
chiral symmetry.
It is found that the properties of the light pseudoscalar 
mesons are reproduced according to the expectations 
from chiral symmetry. 
We then treat the scalar mesons and 
the pseudoscalar mesons 
as the chiral partners. 
In this study, at first we fix the parameters 
so that   the  mass spectrum and 
the decay constants of the pseudoscalar mesons are 
reproduced and 
then we apply those 
for the scalar mesons.

This subject has been studied in other approaches by many authors. 
The most direct approach might be the lattice QCD simulation. 
At present, however, it is difficult to compute the light 
singlet scalar meson. 
There have been studies by using the 
NJL model \cite{HK1994,Dmitra} or the linear sigma model
 \cite{CH1974}. In those studies, the effects of the \UA\  breaking 
 interaction are weak so that the flavor mixing in the $\sigma$ 
 meson is negligible. 
However, recently Takizawa et al. \cite{TNO1997} 
showed that the \UA\  breaking effects must be 
much stronger in order to explain the electromagnetic decays of 
the $\eta$ meson. 
The main aim of this study is to explain the anomalous ordering 
of scalar meson can be explained 
by the  strong \UA\ breaking effects. 
Indeed we will show that the $\sigma-a_{0}$ splitting and significant flavor 
mixing is induced by the strong $U_A(1)$ breaking interaction. 
We will see a large  mixing of $s\sbar$ component in the $\sigma$ meson 
in our approach. 

Other pictures of the light scalar mesons 
proposed so far include 
the multi-quark state or the meson molecule state \cite{BFMNS2001}. 
There may exist the mixing of 
quark-anti quark state with these state. 
However, it is difficult to believe that these multi-quark 
states are the main component of the 
light scalar mesons.
We will show that the scalar meson nonet 
can be explained purely as the quark - anti quark 
state with the \UA\  breaking effects.

In chapter  2, we explain the theoretical background for 
this study. 
In chapter  3, we present the extended NJL model analysis.  
In chapter  4, we present the formulation of the ILA approach. 
In chapter  5, we show our results and give discussions on the mass 
spectrum as well as the mixings.
In chapter  6, conclusions are given.


\chapter{ Theoretical Background}

\section{ Quantum Chromodynamics}

Quantum Chromodynamics (QCD) is the $SU(3)$ nonabelian 
gauge field theory, which describes the 
strong interactions of colored quarks and gluons. 
A quark  comes in 3 colors and gluons come in 
 eight colors. 
Hadrons are color singlet combination of quarks, anti-quarks and gluons. 

The Lagrangian of QCD with the gauge fixing terms is 
\begin{eqnarray}
\L = &&
\psib ({i} \Dsla -m_0  )\psi -\frac{1}{4}F_a^{\mu\nu}F^a_{\mu\nu}
 -\frac{1}{2\xi_0}(\del_{\mu} A^{\mu}_a)^2 \nonumber \\
&& + \del_{\mu}\bar{c}_a \del^{\mu}c_a  + g_0 f_{abe} A_{a}^{\mu}
(\del_{\mu}\bar{c}_b)c_e
\end{eqnarray}
\begin{eqnarray}
F_{\mu\nu}^{a} = \del_{\mu} A_{\nu}^{a} -  \del_{\nu} A_{\mu}^{a} 
-g_0 f^{abc}  A_{\mu}^{b} A_{\nu}^{c}
\end{eqnarray}
\begin{eqnarray}
D_{\mu} = \del_{\mu} + ig_0 \frac{\lambda^a}{2}A_{\mu}^a
\end{eqnarray}
where  $g_0$ is the QCD coupling constant, and the $f_{abc}$ is the 
structure constant of the $SU(3)$ algebra. 
The $\psi_j$ represent the quark fields in which the $j=1,2,3$ is the 
color label. 
The $A_{\mu}^a$ represent the gluon fields with $a=1,..., 8$. 
The quantities $c_a$ are called ghost fields which are 
introduced by in gauge fixing procedure. 
The ghost occurs only in loops, 
and never appears as asymptotic state. 
The $\xi_0$ is the gauge fixing parameter. The limit $\xi_0 \to \infty$ 
defines the Landau  gauge and we use this gauge 
in the improved ladder approximation 
later. 

The most characteristic property of QCD, which supports that the theory 
describing the dynamics of the hadron 
 is QCD, 
is the asymptotic freedom. 
From  the experiments of the deep inelastic scattering 
of the lepton from the hadron, 
it was found that the quarks are almost free in high momentum transfer region. 
This property is called asymptotic freedom 
and it was shown that only the nonabelian gauge field theory 
exhibits it. 
Then the extra symmetry of the nonabelian gauge field 
theory is identified with color symmetry which is 
suggested by the quark model. 
This completes  QCD. 

To see the asymptotic freedom nature of QCD 
we show the running coupling constant 
$\alpha_s(q^2)  \equiv  g^2_R(q^2)/4\pi$.  
It is obtained by the one-loop renormalization group calculation 
as 
\begin{eqnarray}
\alpha_s(q^2) = \frac{12\pi}{(33-2N_f)\ln(q^2/\Lambda^2)} + \cdots
\end{eqnarray}
where $N_f$ is the number of quark flavors with mass less than $\sqrt{q^2}$. 
The $\alpha_s(q^2)$ is small at large $q^2$. 
Consequently in the high energy region, the perturbative 
analysis of QCD makes sense and it is applied to the 
deep inelastic scattering, $e^+e^-$ annihilation process, 
decays of heavy quarkonia and so on. 

In contrast, the $\alpha_s(q^2)$ becomes 
large as $q^2$ decreases and then the perturbation approaches become
poor approximations. 
In this region, the nonperturbative effects, 
especially the confinement and the dynamical chiral 
symmetry breaking, are important. 
The confinement means that only colorless states are 
physically realized and hence quarks cannot be 
observed in isolated state. 
The dynamical chiral symmetry breaking 
is explained in the next section. 
Because the complexity of low energy QCD, 
various approaches have been tried such as 
to solve QCD on the discretized space time numerically (lattice-gauge theory), 
construct effective theories and models 
(th potential quark model, the bag model, the Skyrme model, the NJL model), 
or to consider some limit as  $N_C \to \infty$. 

\section{The Chiral Symmetry Breaking}

If one supposes that all the masses of $N_f$ quarks are zero, 
the QCD Lagrangian satisfies 
${U}_{fL}(N_f) \times {U}_{fR}(N_f) 
 = U_V(1) \times U_A(1) \times {SU}_{fL}(N_f) \times {SU}_{fR}(N_f)$ 
 symmetries, ie.,   
the invariance under the transformations 
\begin{eqnarray}
\psi_L \equiv \frac{1}{2}(1-\g5)\psi  &\to& \psi_L' =  
\exp(i\theta_L^a\frac{\lambda^a_f}{2})\psi_L \qquad (a = 0,...,N_f^2-1)\\
\psi_R \equiv \frac{1}{2}(1+\g5)\psi  &\to& \psi_R' =  
\exp(i\theta_R^a\frac{\lambda^a_f}{2})\psi_R \qquad (a = 0,...,N_f^2-1). 
\end{eqnarray}

Although the quark mass term breaks those symmetries explicitly, 
they remain approximate symmetries of QCD for light flavors i.e., 
$u, d$ and $s$. 

Among these symmetries, some of them are broken in the vacuum 
state. 
The $U_A(1)$ symmetry is broken explicitly by the anomaly 
and we explain it in the next section.

We here concentrate on the 
${SU}_{fL}(3) \times {SU}_{fR}(3)$ 
chiral symmetry, which is broken spontaneously in the vacuum. 
Then the symmetries of the QCD vacuum are 
$U_V(1) \times {SU}_{f}(3)$. 
In this section, we explain the chiral symmetry and its breaking in the 
case of $N_f = 2$ for simplicity. 
At first, we show it using two simple models, 
the linear sigma model and the two-flavor Nambu-Jona-Lasinio (NJL) model, 
which have the $SU(2)_L \times SU(2)_R$ chiral symmetry.
Then we explain 
the chiral symmetry of QCD.

The Lagrangian of the linear sigma model is 
\begin{eqnarray}
\L = &&
\psib \i \dsla \psi 
+ \frac{1}{2}\del_{\mu}{\Vec{\pi}} \del^{\mu}{\Vec{\pi}}
+ \frac{1}{2}\del_{\mu}\sigma \del^{\mu}\sigma 
-g\psib (\sigma - i {\Vec{\tau} \cdot \Vec{\pi}} \g5)\psi
\nonumber \\
&&
+ \frac{\mu^2}{2}(\sigma^2  + \Vec{\pi}^2)
- \frac{\lambda}{4}(\sigma^2  + \Vec{\pi}^2)^2
\end{eqnarray}
where $\psi$ is the doublet nucleon $\psi = \ ^T(p \ n)$ and 
$\Vec{\pi}$ is the triplet of the pion fields  
$\Vec{\pi} = \ ^T(\pi^1 \ \pi^2 \ \pi^3)$.  
This Lagrangian is invariant under the 
${SU}_{fL}(2) \times {SU}_{fR}(2)$ chiral transeformation, 
\begin{eqnarray}
&&\psi_{L,R} \to {\psi'}_{L,R} = U_{L,R}\psi_{L,R}
\qquad 
\Sigma \equiv (\sigma + i \Vec{\tau}\cdot\Vec{\pi})
\to \Sigma' = U_L \Sigma U_R^{\dagger} \\
&&U_{L,R} = \exp(i\theta^a_{L,R} \frac{\tau^a}{2}). 
\end{eqnarray}
It should be noted that this symmetry is isomorphic to 
$O(4)$ symmetry, which is represented by the 4-dimensional 
$(\sigma, \Vec{\pi})$ space. 
When 
$\mu^2$ is 
positive. 
the potential energy 
\begin{eqnarray}
V(\sigma, \Vec{\pi}) = 
 \frac{\mu^2}{2}(\sigma^2  + \Vec{\pi}^2)
- \frac{\lambda}{4}(\sigma^2  + \Vec{\pi}^2)^2
\end{eqnarray}
forms a 4-dimensional  wine bottle shape. 

The classical minimum  of $V(\sigma, \Vec{\pi})$ gives  the 
set of degenerate ground states 
satisfying 
\begin{eqnarray}
\sigma^2 + \Vec{\pi}^2  = \frac{\mu^2}{\lambda}. 
\end{eqnarray}
Let us choose a particular ground state, 
\begin{eqnarray}
\label{Eq:sigmaGround}
\la \sigma \ra = \sqrt{\frac{\mu^2}{\lambda}} \equiv v, 
\quad 
\la \Vec{\pi} \ra = 0
\end{eqnarray}
Then one sees that the chiral $SU(2)_{fL} \times SU(2)_{fR}$ symmetry is 
broken spontaneously by the ground state $(\ref{Eq:sigmaGround})$. 
Defining $\tilde{\sigma} = \sigma - v$, 
we then rewrite the  Lagrangian as 
\begin{eqnarray}
\L  =&& \psi(i \dsla - gv ) \psi 
+\frac{1}{2} [\del_{\mu}\tilde{\sigma}\del^{\mu}\tilde{\sigma} 
-2\mu^2 \tilde{\sigma}^2] 
+ \frac{1}{2}\del_{\mu}\Vec{\pi}\cdot \del^{\mu}\Vec{\pi}
\nonumber \\
&&
-g\psib (\tilde{\sigma} - i {\Vec{\tau} \cdot \Vec{\pi}} \g5)\psi 
+ \lambda v \tilde{\sigma}(\sigma^2  + \Vec{\pi}^2)
- \frac{\lambda}{4}[(\tilde{\sigma}^2  + \Vec{\pi}^2)^2-v^4]. 
\end{eqnarray}
This Lagrangian is not any more symmetric under the rotation in the 
$(\tilde{\sigma},\Vec{\pi})$ space. 
One sees that the pion becomes massless, while the scalar meson 
$\tilde{\sigma}$ acquires a mass $m_{\tilde{\sigma}} = \sqrt{2\mu^2}$, 
and also the fermion gets a mass $M=gv$. 

The appearance of the massless pion is a concrete example of the 
Goldstone theorem: 
if a theory has a continuous symmetry of the Lagrangian 
is spontaneously broken, there must be a massless boson, 
which is called Nambu-Goldstone (NG) boson. 
In this case, the pion is the NG boson. 
The order parameter of the chiral symmetry breaking 
is $v = \la \sigma \ra$. In this model,  the pion decay constant 
$f_{\pi}$ defined by 
\begin{eqnarray}
\la 0 | A_{\mu}^i | \pi^j(\Vec{p}) \ra = i f_{\pi} p_{\mu}\delta^{ij}
\end{eqnarray}
coincides  with $v$ in the tree level.

Next we show the two flavor and $N_C$ color 
NJL model. 
This model is written in terms of  quark degrees of 
freedom as 
\begin{eqnarray}
\L = \psi(i\dsla )\psi  
+ \frac{1}{2}g_s \sum_{i=0}^{3} \left[ (\psib \tau^i \psi)^2 
+ (\psib i\g5 \tau^i \psi)^2 \right]. 
\end{eqnarray}
where $\psi = \ ^T(u \ d)$. 
%
%
This Lagrangian is invariant under the $U(2)_{fL} \times U(2)_{fR}$ chiral 
transformation 
\begin{eqnarray}
&&\psi_{L,R} \to {\psi'}_{L,R} = \exp(i\theta_{L,R}^i \frac{\tau^i}{2})
\psi_{L,R} \quad i = 0,...,3. 
\end{eqnarray}
In this case, the order parameter of the chiral symmetry 
breaking is the value of the composite  operator $\la \psib \psi\ra$ 
and then the symmetry is dynamically broken. 

We analyze this model in the leading order of the $1/N_C$ expansion. 
Using the mean field approximation, the Lagrangian is rewritten as 
\begin{eqnarray}
\label{Eq:NJL1}
&&\L _{\rm MFA} 
= \psib i\dsla q + M \psib \psi \\
\label{Eq:NJL2}
&&M \equiv - \frac{g_s}{N_f} \la \psib \psi \ra 
= \frac{g_s}{N} \lim_{x \to 0}\tr[S_F(x)] \\.
\end{eqnarray}
Eqs. $(\ref{Eq:NJL1})$ and $(\ref{Eq:NJL2})$ lead 
to the following equation for $M$: 
\begin{eqnarray}
\label{NJLSD}
M = 4i g_s N_C \int \frac{d^4p}{(2\pi)^4} \frac{M}{p^2 - M^2}
\end{eqnarray}
Rotating the integral into Euclidean space $(p^0 \to ip_4)$, 
we find the equation: 
\begin{eqnarray}
M = \frac{g_sN_C}{4\pi^2} \int_0^{\Lambda^2}
dp^2 \frac{p^2M}{p^2+M^2}
\end{eqnarray}
where $\Lambda$ is the ultraviolet cutoff. 
After the integral, we obtain 
\begin{eqnarray}
&&M(\kappa - 1)  = \kappa \frac{M^3}{\Lambda^2} \ln \frac{\Lambda^2 +M^2}{M^2}
\\
&&\kappa \equiv \frac{g_sN_C\Lambda^2}{4\pi^2}. 
\end{eqnarray}
There is always the trivial solution, $M = 0$, to this equation, and 
when $\kappa > 1$, there is also a nontrivial one with $M \neq 0$. 
In the case of the $M \neq 0$, the value of $\la \psib \psi \ra$ 
is not zero and it shows that the chiral $U_{fL}(2) \times U_{fR}(2)$ 
is dynamically broken to the $U_{fv}(2)$.  
The quarks acquire the dynamical mass $M$. 

According to the dynamical symmetry breaking, 
there should be the four NG bosons. 
To confirm this, we consider the Bethe-Salpeter (BS) equations 
for the pseudoscalar mesons and the scalar mesons. 
The BS equation in the present model in Euclid space is written as 
\begin{eqnarray}
&&{S_F}^{-1}_{n,n_1}(q+\frac{P}{2}) \chi_{n_1m_1} (q,P) {S_F}_{m_1m}^{-1}(q-\frac{P}{2})
= \int \frac{d^4k}{(2\pi)^4}
K_{mn,n_2m_2}(q,k;P)\chi_{n_2,m_2}(q;P)
\nonumber \\
&& K_{mn,n_2m_2} = i\frac{g_s}{2} 
\left[
(\tau^j)_{nm}(\tau^j)_{m_2n_2} + (i\g5\tau^j)_{nm}(i\g5\tau^j)_{m_2n_2}
\right]
\end{eqnarray}
where $\chi$ is the BS amplitude defined as 
\begin{eqnarray}
\chi_{nm}(q;P) = e^{-iP\frac{x+y}{2}}\int d^4(x-y)
\la 0 | T \psib_n(x) \psi_m(y) | P \ra e^{iq(x-y)}. 
\end{eqnarray}
$|P \rangle$ is the bound state with momentum $P$. 
The BS amplitude for the pseudoscalar meson, ${\chi^{(P)}}_i$, 
and the BS amplitude for the scalar meson, ${\chi^{(S)}}_i$, 
can be written in terms of 
four scalar amplitudes as 
\begin{eqnarray}
&&{\chi^{(P)}_j}_{nm}(q;P) =  \Vec{1}_{C} \frac{\tau_{j}}{2}
\left[ \phi^{(P)}_S(q;P) + \phi^{(P)}_P(q;P) \qsla + \phi^{(P)}_Q(q;P)\Psla
+\phi^{(P)}_T(q;P) \frac{1}{2}(\Psla \qsla - \qsla \Psla)\right]\g5
\nonumber \\
&&{\chi_j^{(S)}}_{nm}(q;P) =  \Vec{1}_{C} \frac{\tau_{j}}{2}
\left[ \phi^{(S)}_S(q;P) + \phi^{(S)}_P(q;P) \qsla + \phi^{(S)}_Q(q;P)\Psla
+\phi^{(S)}_T(q;P) \frac{1}{2}(\Psla \qsla - \qsla \Psla)\right]
\nonumber
\end{eqnarray}
where $\tau^a$ is the Pauli matrix which denotes 
the flavor structure of the meson state. 
In the chiral limit and in the leading order of the $1/N_C$ 
expansion, the BS equations becomes extremely simple. 
Furthermore we takes only the first order in $P^2/\Lambda^2$, 
then we obtain the following BS equation: 
\begin{eqnarray}
\label{NJLBS_p}
\left(q^2 + \frac{P^2}{2} \right) \phi_S^{(P)}(q^2,P^2) 
&=& \frac{\kappa}{\Lambda^2}\int^{\Lambda^2}dk^2 k^2 \phi_S^{(P)}(k^2,P^2)\\
\label{NJLBS_s}
\left(q^2 + \frac{P^2}{2} + 4M \right) \phi_S^{(S)}(q^2,P^2) 
&=& \frac{\kappa}{\Lambda^2}\int^{\Lambda^2}dk^2 k^2 \phi_S^{(S)}(k^2,P^2). 
\end{eqnarray}
These equations are satisfied for each flavor $j$.  

If one identifies the amputated function 
$\tilde{\chi}(q^2) = (M^2 - q^2) \chi$ and 
substitute the $\frac{P^2}{2} \to M^2$,  
Eq. $(\ref{NJLBS_p})$  coincides with Eq. ($\ref{NJLSD}$). 
Then we can easily solve it. 

The spectrum of $M_{P}^2 = -P^2$ of pseudoscalar bound states and the 
spectrum of $M_{S}^2 = -P^2$ of scalar bound states 
are determined from the following equations 
\begin{eqnarray}
&&\kappa - 1 = \kappa \frac{M^2 - \frac{M_P^2}{2}}{\Lambda^2}
\ln \left(\frac{\Lambda^2 + M^2 -\frac{M_P^2}{2}}{M^2 - \frac{M^2_S}{2}}\right)
\\
&&\kappa - 1 = \kappa \frac{M^2 - \frac{M_S^2}{2}}{\Lambda^2}
\ln \left(\frac{\Lambda^2 + 3M^2 -\frac{M_S^2}{2}}{3M^2 - \frac{M^2_S}{2}}\right). 
\end{eqnarray}
We obtain for $M \neq 0$ case 
\begin{eqnarray}
&&M_P = 0\\
&&M_S = 2M. 
\end{eqnarray}
The four pseudoscalar mesons are massless and they are the NG bosons 
for the chiral symmetry breaking $U_{fL}(2) \times U_{fR}(2) \to U_{fV}(2)$.

In the case of QCD, 
if all the $u,d,s$ quarks are massless, which is called the chiral limit, 
the chiral ${SU}_{fL}(3) \times {SU}_{fR}(3)$ symmetry is exact. 
Although it is broken by the quark mass 
term explicitly, since those masses are much smaller than 
the typical hadron masses $(\sim 700[\MeV/c^2] - 1 [\GeV/c^2] )$, 
this symmetry is approximately satisfied. 
By considering the chiral symmetry and 
its dynamical breaking, 
many properties of light hadrons  are  
understood. 
Especially, the fact that the pion is extremely lighter than the other 
mesons  is explained as the pion is the  Nambu-Goldstone (NG) boson 
 accompanying the dynamical breaking of the chiral symmetry. 
Practically, in the QCD vacuum the chiral symmetry is  
dynamically broken and the residual symmetry is $SU_V(3)$. 
In that time, eight symmetry is broken and then 
there are eight NG bosons. 
The light pseudoscalar octet mesons excluding $\eta'$ are the NG boson. 
The $\eta$ and $\eta'$ mesons are the 
mixed state of the octet $NG$ boson state and 
the singlet state which is not the NG boson state. 
This non-zero value of the quark condensate can not be obtained by 
the perturbative calculation. 
It is caused by the nonperturbative effects. 
The ladder approximation which we use in this study 
is the convenient way to obtain 
the non-zero value of quark condensate.  

Additionally, in the chiral symmetry broken vacuum 
the effective quark mass is generated.

\section{$U_A(1)$ Symmetry Breaking}

For understanding the properties of the light 
hadrons, it is necessary to 
consider the \UA\ symmetry, which is
expected to be broken by anomaly. 
The QCD Lagrangian is symmetric under  the $U_A(1)$ 
transformation. 
However, Weinberg showed that the mass of $\eta'$ should be less
than $\sqrt{3} m_{\pi}$ if \UA\ symmetry were not
explicitly broken.\cite{Weinberg1975}
The experimental value of the $eta'$ mass is $957.78 \pm 0.14 [\MeV]$ 
and that of the pion mass is $\sim 140 [\MeV]$.
Thus the \UA\ symmetry must be broken by the anomaly. 
Later, 't Hooft pointed out the relation between 
\UA\ anomaly and topological gluon configurations of QCD and 
showed that the interaction of light quarks and instantons breaks 
the \UA\ symmetry.\cite{tHooft1976}

The instanton is the classical topological solution
 of the gauge field in the 4-dimensinal Euclid space. 
Then it is self-dual $F^a_{\mu\nu} =\tilde{F^a_{\mu\nu}}$ 
where 
$\tilde{F}_a^{\mu\nu} \equiv \frac{1}{2}\epsilon ^{\mu\nu\alpha\beta} F_{\alpha\beta}^a$. 
The BPST \cite{BPST} instanton solution in the singular gauge is 
given by 
\begin{eqnarray}
&&
A^a_{\mu}(x) = 2\frac{x_{\nu}}{x^2}
\frac{\bar{\eta}_{a\mu\nu}\rho^2}{x^2+\rho^2} \\
&&\bar{\eta_{a\mu\nu}}=
\left\{
\begin{array}{cl}
\epsilon_{a\mu\nu} & \mu,\nu=1,2,3 \\
-\delta_{a\mu} & \nu = 4 \\
\delta_{a\nu} & \mu = 4 \\
\end{array}
\right.
\end{eqnarray}
where $\rho$ is the instanton size. 
This solution has the non zero value 
\begin{eqnarray}
q = \int d^4x \frac{g^2}{32\pi^2}
F^a_{\mu\nu}\tilde{F}_a^{\mu\nu}
\end{eqnarray}
This integral is the value related to the topology which called as 
Pontryagin index q. 
In the Minkowski space, the instanton gives the transition amplitude 
from the vacuum of topological index Q to the vacuum of topological 
index Q+q. 

The QCD vacuum may be  considered as the multi-instanton state. 
The dynamics of instantons in the multi-instanton vacuum has
been studied by many authors, either in the models or in the
lattice QCD approach, and the widely accepted picture is that
the QCD vacuum consists of small instantons of the size about
1/3 fm with the density of 1 instanton (or anti-instanton) per
fm$^{4}$.\cite{Instanton}

If the vacuum consists of instanton configurations, 
light quarks are affected by the instantons. 
Especially for a massless quark 
there exists  a zero mode which satisfies 
\begin{eqnarray}
(-i)\Dsla \ \psi_0 = 0. 
\end{eqnarray}
The solution is given by 
\begin{eqnarray}
{\psi_0}(x) = -\frac{\rho}{\pi}\frac{x_{\mu} \gamma_{\mu}}{\sqrt{x^2}}
\frac{1}{(x^2+\rho^2)^{3/2}}
\left(
\begin{array}{c}
\phi \\
-\phi \\
\end{array}
\right)
\end{eqnarray}
where $\phi$ is the hedgehog structure for the $SU(2)$ color 
and spin. 
This solution satisfies the $\g5 \psi_0 = \psi_0$ and then the 
instanton induces 
transition from the left hand $\psi_{0L}$ to the right hand  $\psi_{0R}$ 
may occur. 
Since the zero mode exists in each flavor, 
the Green's function 
the  $N_f$ quark and $N_f$ antiquark 
 $\la \Pi_{N_f}\psib_R\psi_L \ra$, which is not $U_A(1)$ 
invariant, does  not vanish. 

This effect can be represented  by an low energy 
effective interaction with $2N_f$ quark vertex. 
In the case of $N_f = 3$, it is given by a 
six-quark interaction as  
\begin{eqnarray}
{\cal L}_6 & = & G_D \left\{ \, {\rm det} \left[ \bar \psi_i (1 - \gamma_5) 
\psi_j \right] + {\rm det}  \left[ \bar \psi_i (1 + \gamma_5) \psi_j 
\right] \, \right\} \, 
\end{eqnarray}
which is called the Kobayashi-Maskawa-'t Hooft (KMT) interaction. 
This interaction is not $U_A(1)$ symmetric and is 
$SU(3)_{fR} \times SU(3)_{fR}$ symmetric. 
By substituting the one $\bar{\psi}\psi$ by the condensate or mass term 
, the four-quark flavor mixing interaction is produced. 

The effects of the KMT interaction on  the scalar $q\bar{q}$ state 
and the pseudoscalar $q\bar{q}$ state are summarized 
in Table. $\ref{TBL:KMTforce}$.
From this table, one sees in how the KMT interaction 
induces the $\eta - \eta'$ mass difference. 
which can be understood by flavor mixing in the $I=0$ 
${1\over \sqrt{2}} \left( u\ubar+d\dbar \right)$ 
and $s\sbar$. 
Without the flavor mixing, $\eta = {1\over \sqrt{2}} \left( u\ubar+d\dbar \right)$ 
and $\eta' = s\sbar$ would form 
mass eigenstates, and thus the ideal mixing is achieved. 
This is natural if the Okubo-Zweig-Iizuka (OZI) rule applies.
However, the OZI rule is known to be significantly broken in 
the pseudoscalar mesons.
By the flavor mixing effect of the  KMT interaction, 
these states are mixed. 
Then 
the $\eta$ 
approaches to the flavor octet state 
${1 \over \sqrt{6}} \left( u\ubar+d\dbar-2s\sbar \right)$ 
and $\eta'$ approaches to the flavor singlet state 
${1\over \sqrt{6}} \left( u\ubar+d\dbar+s\sbar \right)$. 
For instance, according to the 
Nambu-Jona-Lasinio(NJL) model,  the electromagnetic $\eta$ decay 
processes suggest  that the mixing of 
${1\over \sqrt{2}} \left( u\ubar+d\dbar \right)$ 
and $s\sbar$ is indeed strong so that the $\eta$ meson is close to  
the pure octet state.\cite{TNO1997}
Then in $\eta$, the KMT interaction 
works as the attractive force which competes with 
the effects of the increase of the $s\bar{s}$ 
component. 
On the contrary, 
in $\eta'$, the KMT interaction 
works as the repulsive  force which competes with 
the effects of the decrease  of the $s\bar{s}$ 
component.
According to the improved ladder approximation approach which we will show 
later, 
as the KMT interaction increases, 
$\eta$ mass does not change after slightly increasing 
and $\eta'$ mass increase monotonically. 
Consequently,  the $\eta - \eta'$ mass difference is explained 
by the $U_A(1)$ breaking effects of the KMT interaction. 

In this study, we investigate the role of 
this $U_A(1)$ breaking effect in the 
light scalar nonet meson mass spectrum.

%
\begin{table}[btp]
\begin{center}
\begin{tabular}{|c|cc|}  \hline
flavor &pseudoscalar & scalar \\\hline
octet & attractive   & repulsive   \\
singlet & repulsive  & attractive  \\
\hline
\end{tabular} 
\end{center}  
 \caption{The effects of the Kobayashi-Maskawa-t'Hooft interaction 
 on the pseudoscalar $q\bar{q}$ state and the scalar $q\bar{q}$ state.
 }
\label{TBL:KMTforce}
\end{table}
%

\section{The Improved  Ladder Approximation of QCD}
The improved ladder approximation of QCD is based on the approximation 
in the Schwinger-Dyson (SD) and Bethe-Salpeter (BS) equation of QCD. 
The gluon degrees of freedom are integrated out and are represented 
by the quark-quark interaction kernels. Both SD equation and 
BS equation are restricted to the ladder diagrams of gluon exchange 
between quark lines and the gluon polarization and the 
vertex correction are taken into account through 
the running coupling constant approximately.


The origin of this approach is 
the Higashijima-Miransky approach \cite{Higashijima} for the 
Swinger-Dyson (SD) equation for the quark propagator, 
which 
keeps  consistency  with asymptotic freedom 
by using  the running coupling constant of 
one loop renormalization group calculation. 
Since that running coupling constant 
diverges at low momentum, 
an infrared cutoff  are introduced. 
Under the rainbow approximation 
the SD equation in the Landau gauge 
shows the D$\chi$SB and gives 
the running quark mass consistent with the 
renormalization group 
at high momentum. 

%
%

Later, Aoki et al. \cite{ILA} applied the same 
idea to the BS equation and shows that the 
combination of the rainbow SD and the ladder BS equation is 
consistent with chiral symmetry. It was shown that in the 
chiral limit, the pseudoscalar meson obtained as solutions 
of the BS equation have the expected properties consistent 
with chiral symmetry.

The shortcomings of this approach may be summarized 
in the following four points. 
The first one is that the infrared cutoff can not be 
deduced from QCD.  
Although  the strength of the 
D$\chi$SB depends on it strongly,  
it should be given  
phenomenologically.
The second one is that results are gauge dependet. 
In the ILA, the Landau gauge 
is employed at the gluon propagator. 
To remove the gauge dependence,  
the higher order may be necessary. \cite{AokiNPRG} 
The third one is that in ILA a large $\lamQ$ is 
needed to reproduce the 
appropriate magnitude of the chiral symmetry 
breaking. 
It may indicate  that the chiral symmetry breaking 
can not be explained only by the 
ladder approximation of the one gluon exchange. 
The last one is that the axial Ward identity of QCD 
is not satisfied 
by the non-local coupling constant of the 
gluon exchange interaction. 
However, by using the modified axial vector current 
the axial Ward identity can be satisfied. \cite{nairon1_2}
In this study, this modified axial vector current is used. 

A realistic model based on the ILA has been studied by 
Naito et al. \cite{NNTYO2000}
They introduced finite quark mass and further 
$U_A(1)$ breaking interaction in the 
interaction kernel and calculated the pseudoscalar 
spectrum.


\chapter{Extended Nambu-Jona-Lasinio model}

In order to clarify the roles of the breaking interaction 
in the scalar meson spectrum, we first consider a simple 
model and analyze the pseudoscalar and scalar mesons. 
One of the simplest models which incorporate chiral 
symmetry and its dynamical  breaking 
is the Nambu-Jona-Lasinio (NJL) model, 
which is extended by an additional $U_A(1)$ 
breaking interaction.  
The $U_A(1)$ breaking is considered to be 
caused by the interaction of 
quarks and instantons in the vacuum. 
Here the $U_A(1)$ breaking interaction is represented by the 
six-quark interaction term called 
Kobayashi-Maskawa-t' Hooft (KMT) interaction. 
\section{Formulation}
We work with the following NJL model lagrangian density extended to 
three-flavor case:
\bea
{\cal L} & = & {\cal L}_0 + {\cal L}_4 + {\cal L}_6 , \label{njl1} \\
{\cal L}_0 & = & \bar \psi \,\left( i \partial_\mu \gamma^\mu - \hat m 
\right) \, \psi \, ,
\label{njl2} \\
{\cal L}_4 & = & {G_S \over 2} \sum_{a=0}^8 \, \left[\, \left( 
\bar \psi \lambda^a \psi \right)^2 + \left( \bar \psi \lambda^a i \gamma_5 
\psi \right)^2 \, \right] \, ,
\label{njl3} \\
{\cal L}_6 & = & G_D \left\{ \, {\rm det} \left[ \bar \psi_i (1 - \gamma_5) 
\psi_j \right] + {\rm det}  \left[ \bar \psi_i (1 + \gamma_5) \psi_j 
\right] \, \right\} \, .
\label{njl4}
\eea
Here the quark field $\psi$ is a column vector in color, flavor and Dirac 
spaces and $\lambda^a (a=0\ldots 8)$ is the Gell-Mann matrices for 
the flavor $U(3)$. 
The free Dirac lagrangian ${\cal L}_0$ incorporates the current quark mass 
matrix $\hat m = {\rm diag}(m_u, m_d, m_s)$ which breaks the chiral 
$U_L(3) \times U_R(3)$ invariance explicitly. ${\cal L}_4$ is a QCD 
motivated four-fermion interaction, which is chiral $U_L(3) \times U_R(3)$
invariant.  The Kobayashi-Maskawa-'t Hooft determinant ${\cal L}_6$ 
represents the $U_A(1)$ anomaly.  
It is a $3 \times 3$ determinant with respect to flavor with 
$i,j = {\rm u,d,s}$.   

Quark condensates and constituent quark masses are self-consistently 
determinded by the gap equations in the mean field approximation,
\bea
M_u & = & m_u - 2G_S \la \ubar u \ra - 
  2 G_D \la \dbar d \ra \la \sbar s \ra \, , \nonumber \\
M_d & = & m_d - 2G_S \la \dbar d \ra - 
  2 G_D \la \sbar s \ra \la \ubar u \ra \, , \nonumber \\
M_s & = & m_s - 2G_S \la \sbar s \ra - 
  2 G_D \la \ubar u \ra \la \dbar d \ra \, , \label{gap}
\eea
with 
\bea
\la \qbar q \ra & = & - {\rm Tr}^{(c,D)} \left[ iS_F^q (x = 0) \right] 
\nonumber \\
& = & - \int^\Lambda \frac{d^4p}{(2\pi)^4} {\rm Tr}^{(c,D)}
\left[ \frac{i}{p_\mu \gamma^\mu - M_q + i\epsilon} \right] \, .
\label{condensate}
\eea
Here the covariant cutoff $\Lambda$ is introduced to regularize the
divergent integral and Tr$^{(c,D)}$ means trace in color and Dirac spaces.

The scalar channel quark-antiquark scattering amplitudes
\begin{equation}
\la p_3 , \bar p_4 ; {\rm out} \right. \left| p_1 , \bar p_2 ; {\rm in} \ra 
 =  (2 \pi)^4 \delta^4(p_3 + p_4 - p_1 - p_2) {\cal T}_{q \bar q} 
\end{equation}
are then calculated in the ladder approximation. 
We assume that $m_u = m_d$ so that the isospin is exact.
In the $\sigma$ and $f_0$ channel, 
the explicit expression is 
\begin{equation}
{\cal T}_{q \bar q} = -
\left(
\begin{array}{c} 
\bar u(p_3) \lambda^8 v(p_4) \\
\bar u(p_3) \lambda^0 v(p_4) 
\end{array} 
\right)^T \,
\left(
\begin{array}{cc}
A(q^2) & B(q^2) \\
B(q^2) & C(q^2) \\
\end{array}
\right) \,
\left(
\begin{array}{c}
\bar v(p_2) \lambda^8 u(p_1) \\
\bar v(p_2) \lambda^0 u(p_1)
\end{array}
\right) \, , \label{qas1}
\end{equation}
with 
\begin{eqnarray}
A(q^2) & = & \frac{2}{{\rm det}{\bf D}(q^2)} 
\left\{ 2 ( G_0 G_8 - G_m G_m ) I^0 (q^2) - G_8 \right\} \, , \label{qas2} \\
B(q^2) & = & \frac{2}{{\rm det}{\bf D}(q^2)}
\left\{- 2 ( G_0 G_8 - G_m G_m ) I^m (q^2) - G_m \right\} \, , \label{qas3} \\
C(q^2) & = & \frac{2}{{\rm det}{\bf D}(q^2)}
\left\{ 2 ( G_0 G_8 - G_m G_m ) I^8 (q^2) - G_0 \right\} \, , \label{qas4} 
\end{eqnarray}
and 
\bea
G_0 & = & \frac{1}{2} G_S - \frac{1}{3} ( 2 \langle \bar uu \rangle + 
\langle \bar ss \rangle ) G_D \, , \\
G_8 & = & \frac{1}{2} G_S - \frac{1}{6} ( \langle \bar ss \rangle - 4  
\langle \bar uu \rangle ) G_D \, , \\
G_m & = & - \frac{1}{3 \sqrt{2}} ( \langle \bar ss \rangle - 
\langle \bar uu \rangle ) G_D \, .
\eea
The quark-antiquark bubble integrals are defined by
\begin{eqnarray}
I^0(q^2) & = & i \int^{\Lambda} \frac{d^4p}{(2 \pi)^4} {\rm Tr}^{(c,f,D)}
\left[ S_F(p) \lambda^0 S_F(p+q) \lambda^0 \right]
\, , \label{int1} \\
I^8(q^2) & = & i \int^{\Lambda} \frac{d^4p}{(2 \pi)^4} {\rm Tr}^{(c,f,D)}
\left[ S_F(p) \lambda^8 S_F(p+q) \lambda^8 \right]
\, , \label{int2} \\
I^m(q^2) & = & i \int^{\Lambda} \frac{d^4p}{(2 \pi)^4} {\rm Tr}^{(c,f,D)}
\left[ S_F(p) \lambda^0 S_F(p+q) \lambda^8 \right]
\, , \label{int3} 
\end{eqnarray}
with $q = p_1 + p_2$.  The $2 \times 2$ matrix ${\bf D}$ is given by
\begin{equation}
{\bf D}(q^2) = 
\left( 
\begin{array}{cc}
D_{11}(q^2) & D_{12}(q^2) \\
D_{21}(q^2) & D_{22}(q^2) 
\end{array}
\right) \, , \label{mat}
\end{equation}
with
\begin{eqnarray}
D_{11}(q^2) & = & 2 G_0 I^0(q^2) + 2 G_m I^m(q^2) - 1 \, , \label{mat11}\\
D_{12}(q^2) & = & 2 G_0 I^m(q^2) + 2 G_m I^8(q^2) \label{mat12} \\
D_{21}(q^2) & = & 2 G_8 I^m(q^2) + 2 G_m I^0(q^2) \label{mat21} \\
D_{22}(q^2) & = & 2 G_8 I^8(q^2) + 2 G_m I^m(q^2) - 1 \, . \label{mat22}
\end{eqnarray}
From the pole positions of the scattering amplitude Eq. (\ref{qas1}), the 
$\sigma$-meson mass $m_{\sigma}$ and the $f_0$-meson mass $m_{f_0}$ 
are determined.
\par
    The scattering amplitude Eq. (\ref{qas1}) can be diagonalized by rotation
in the flavor space 
\begin{eqnarray}
{\cal T}_{q \bar q} & = & -
\left(
\begin{array}{c} 
\bar u(p_3) \lambda^8 v(p_4) \\
\bar u(p_3) \lambda^0 v(p_4) 
\end{array} 
\right)^T   {\bf T}_{\theta}^{-1} {\bf T}_{\theta} 
\left(
\begin{array}{cc}
A(q^2) & B(q^2) \\
B(q^2) & C(q^2) \\
\end{array}
\right) {\bf T}^{-1}_{\theta}  \nonumber \\
&& \times {\bf T}_{\theta}  
\left(
\begin{array}{c}
\bar v(p_2) \lambda^8 u(p_1) \\
\bar v(p_2) \lambda^0 u(p_1)
\end{array}
\right) \, , \label{qasm1} \\
& = & -
\left(
\begin{array}{c} 
\bar u(p_3) \lambda^{\sigma} v(p_4) \\
\bar u(p_3) \lambda^{f_0}  v(p_4) 
\end{array} 
\right)^T \, 
\left(
\begin{array}{cc}
D^{\sigma}(q^2) & 0 \\
0 & D^{f_0}(q^2) 
\end{array}
\right) \nonumber \\
&& \times
\left(
\begin{array}{c}
\bar v(p_2) \lambda^{\sigma} u(p_1) \\
\bar v(p_2) \lambda^{f_0} u(p_1)
\end{array}
\right) \, , \label{qasm2}
\end{eqnarray}
with $\lambda^{\sigma} \equiv \cos \theta  \lambda^8 - \sin \theta \lambda^0$,
$\lambda^{f_0} \equiv \sin \theta  \lambda^8 + \cos \theta \lambda^0$ and
\begin{equation}
{\bf T}_{\theta} = \left(
\begin{array}{cc}
\cos \theta & -\sin \theta \\
\sin \theta & \cos \theta 
\end{array}
\right) \, .
\end{equation}
The rotation angle $\theta$ is determined by 
\begin{equation}
\tan 2 \theta = \frac{2 B(q^2)}{C(q^2) - A(q^2)} \, . \label{angle}
\end{equation}
Note that $\theta$ therefore depends on $q^2$.  At $q^2 = m_{\sigma}^2$,  
$\theta$ represents the 
mixing angle of the $\lambda^8$ and $\lambda^0$ components in the 
$\sigma$-meson state.

The \UA\ breaking KMT 6-quark determinat interaction ${\cal L}_6$ contributes
to the scalar $q \qbar$ channel only by the form of the effective 4-quark 
interaction, which is derived from ${\cal L}_6$ by contracting 
a quark-antiquark pair into the quark condensate. The explicit form of 
the effective KMT interaction is 
\bea
{\cal L}_6^{eff} =\left( \frac{-1}{2} \right) G_D \Biggl\{ & &
\left( \frac{-2}{3} \right) \left( 2 \la \ubar u \ra + \la \sbar s \ra \right)
\left[ \left(\bar \psi \lambda^0 \psi \right)^2 
     - \left(\bar \psi \lambda^0 i \gamma_5 \psi \right)^2 \right] 
\nonumber \\
& & + \la \sbar s \ra \sum_{i=1}^3 
\left[ \left(\bar \psi \lambda^i \psi \right)^2 
     - \left(\bar \psi \lambda^i i \gamma_5 \psi \right)^2 \right] 
\nonumber \\
& & + \la \ubar u \ra \sum_{i=4}^7 
\left[ \left(\bar \psi \lambda^i \psi \right)^2 
     - \left(\bar \psi \lambda^i i \gamma_5 \psi \right)^2 \right] 
\nonumber \\
& & + \left( \frac{1}{3} \right) 
\left( 4 \la \ubar u \ra - \la \sbar s \ra \right)
\left[ \left(\bar \psi \lambda^8 \psi \right)^2 
     - \left(\bar \psi \lambda^8 i \gamma_5  \psi \right)^2 \right] 
\nonumber \\
& & + \left( \frac{\sqrt{2}}{3} \right) 
\left( \la \ubar u \ra - \la \sbar s \ra \right) 
\Bigl[ \left(\bar \psi \lambda^0 \psi \right) 
       \left(\bar \psi \lambda^8 \psi  \right) 
+ \left(\bar \psi \lambda^8 \psi \right) 
       \left(\bar \psi \lambda^0 \psi  \right) 
\nonumber \\
& &  - \left(\bar \psi \lambda^0 i \gamma_5 \psi \right) 
       \left(\bar \psi \lambda^8 i \gamma_5 \psi  \right) 
+ \left(\bar \psi \lambda^8 i \gamma_5 \psi \right) 
       \left(\bar \psi \lambda^0 i \gamma_5 \psi  \right) \Bigr] 
\Biggr\} \, . \label{ekmt}
\eea
One can easily figure out from Eq. (\ref{ekmt}) that the \UA\ breaking KMT
interaction gives the attractive force in the flavor singlet scalar $q \qbar$ 
channel. On the other hand, it gives the repulsive force in the 
isospin $I = 1$ ($a_0$) and $I = 1/2$ ($K_0^*$) channels. 
Because of the large strange quark mass, $| \la \sbar s \ra |$ is bigger than 
$| \la \ubar u \ra |$, and therefore, the repulsion 
in the $I = 1$ channel is stronger than that in the $I = 1/2$ channel.
\section{Results}
We show our numerical results and give discussions on the mass 
spectrum as well as the mixings in this section.
As the extended NJL model has been used in the analyses of the pseudoscalar 
mesons, here we have used the model parameters fixed in the study of the 
electromagnetic decays of the $\eta$ meson. Since the $\eta$ meson 
properties depend on the strength of the \UA\
breaking interaction rather sensitively, 
it is reasonable to determine the strength of the
\UA\ breaking interaction from the $\eta$ meson properties.

The parameters of the NJL model are the current quark masses 
$m_{u}=m_{d}$, $m_{s}$, the four-quark coupling constant $G_{S}$, the 
\UA\ breaking KMT six-quark determinant coupling constant $G_{D}$ and the 
covariant cutoff $\Lambda$.  We take $G_{D}$ as a free parameter and study 
scalar meson properties as functions of $G_{D}$.
We use the light current quark masses $m_{u}=m_{d}=8.0$ MeV to reproduce 
$M_u = M_d \simeq 330$ MeV ($\simeq 1/3 M_N$) which is the value
commonly used in the constituent quark model.
The other parameters, $m_{s}$, $G_{S}$ and $\Lambda$, are determined
so as to reproduce the isospin averaged observed masses, $m_{\pi} = 138.0$ 
MeV, $m_{K} = 495.7$ MeV and the pion decay constant $f_{\pi} = 92.4$ MeV.
When we take the different value of $G_{D}$, we go through the fitting 
procedure each time.
\par
    We obtain $m_{s}=193$ MeV, $\Lambda=783$ MeV, $M_{u,d}=325$ MeV,
$M_{s}=529$ MeV and 
$f_K = 97$ MeV, which are almost independent of $G_{D}$.
The quark condensates are also independent of $G_D$ and our results are 
$\langle \bar uu \rangle^{\frac{1}{3}} = -216$ MeV and 
$\langle \bar ss \rangle^{\frac{1}{3}} = -226$ MeV whenever we have fixed 
other model parameters from the observed values of $m_\pi$, $m_K$ and $f_\pi$.
\par
    We define dimensionless parameters,
\bea
G_{D}^{\rm eff} &\equiv& - G_{D} (\Lambda / 2 \pi)^{4} \Lambda N_{c}^{2}
\nonumber\\
G_{S}^{\rm eff} &\equiv& G_{S} (\Lambda / 2 \pi)^{2} N_{c} .
\eea
As reported in Ref.~\cite{TNO1997}, the experimental value of the 
$\eta \to \gamma \gamma$ decay amplitude is reproduced at about 
$G_D^{\rm eff} = 0.7$. The calculated $\eta$-meson mass at 
$G_D^{\rm eff} = 0.7$ is $m_{\eta} = 510$ MeV which is 7\% smaller than 
the observed mass.  
$G_D^{\rm eff} = 0.7$ corresponds to 
$G_{D} \langle \overline{s} s \rangle / G_{S} =0.44$, suggesting that 
the contribution from ${\cal L}_{6}$ to the dynamical mass of 
the up and down quarks is 44\% of that from ${\cal L}_{4}$.
The calculated value of $\Gamma(\eta \to \pi^0 \gamma \gamma)$ is 0.92 eV 
at $G_D^{\rm eff} = 0.7$, which is in good agreement with the experimental 
data: $\Gamma(\eta \to \pi^0 \gamma \gamma) = 0.93 \pm 0.19$ eV.

Before going to present the numerical results for the scalar mesons, let us 
summarize the properties of the scalar mesons in the NJL model.
In the $SU_L(2) \times SU_R(2)$ version of the NJL with no explicit 
symmetry breaking term, the $\sigma$-meson mass can be calculated 
analytically in the mean field + ladder approximation, \ie, 
$m_\sigma = 2 M_u$.  
The $\sigma$ meson is therefore regarded as the lowest bosonic excitation, 
whose mass is twice of the gap energy, associated with chiral symmetry 
breaking.
It should be noticed that there is a cut above $q^2 = 4 M_u^2$ in the complex 
$q^2$-plane of the quark-antiquark scattering T-matrix, 
which corresponds to 
the unphysical decay: $\sigma \to \qbar q$. This is one of the known
shortcomings of the NJL model.  If one introduces a small symmetry breaking 
term, \ie, the current quark mass term, $m_\sigma$ moves up and gets the 
imaginary part corresponding to the $\sigma \to \qbar q$ decay.
\cite{TTKK1990}
The pole position is in the 
second Riemann-sheet of the complex $q^2$-plane,  as is the case of ordinary
resonances. It means that the Argand diagram for the T-matrix makes a circular
resonance shape in the scalar $q \qbar$ channel.%
\footnote{The situation is quite different in the case of the vector meson. 
In the nonrelativistic limit, the scalar meson channel corresponds to the 
p-wave quark-antiquark state whereas the vector meson channel corresponds to 
the s-wave quark-antiquark state. See Ref.~\cite{TKM1991}.}

It should be noted that the physical decay mode of $\sigma$, \ie,
$\sigma\to\pi\pi$ is neither taken into account in the ladder approximation.
As this decay makes the $\sigma$ width significantly large, our result for 
$\sigma$ mass is qualitative rather than quantitative.
Nevertheless, the results shown below show that the scalar mesons in 
the NJL model is realized as the chiral partner of the 
Nambu-Goldstone bosons, and that they give systematic behavior for 
the orders of the masses and the splittings.

\begin{figure}[t]
	\centerline{\epsfxsize= 10cm \epsfbox{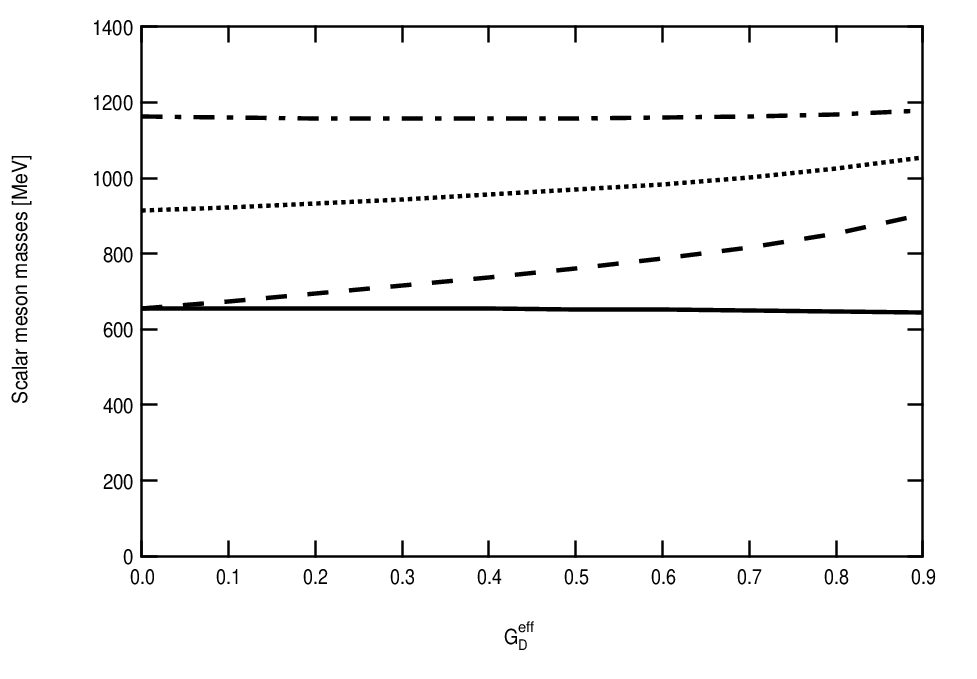}}
\caption{The calculated scalar meson masses as functions of the effective
coupling constant $G_D^{\rm eff}$ of the \UA\ breaking KMT interaction.
The solid, dashed, dotted and dash-dotted lines represent 
$m_\sigma$, $m_{a_0}$, $m_{K_0^*}$ and $m_{f_0}$, respectively.}
\label{fig:mass}
\end{figure}

\begin{figure}[t]
	\centerline{\epsfxsize= 10cm \epsfbox{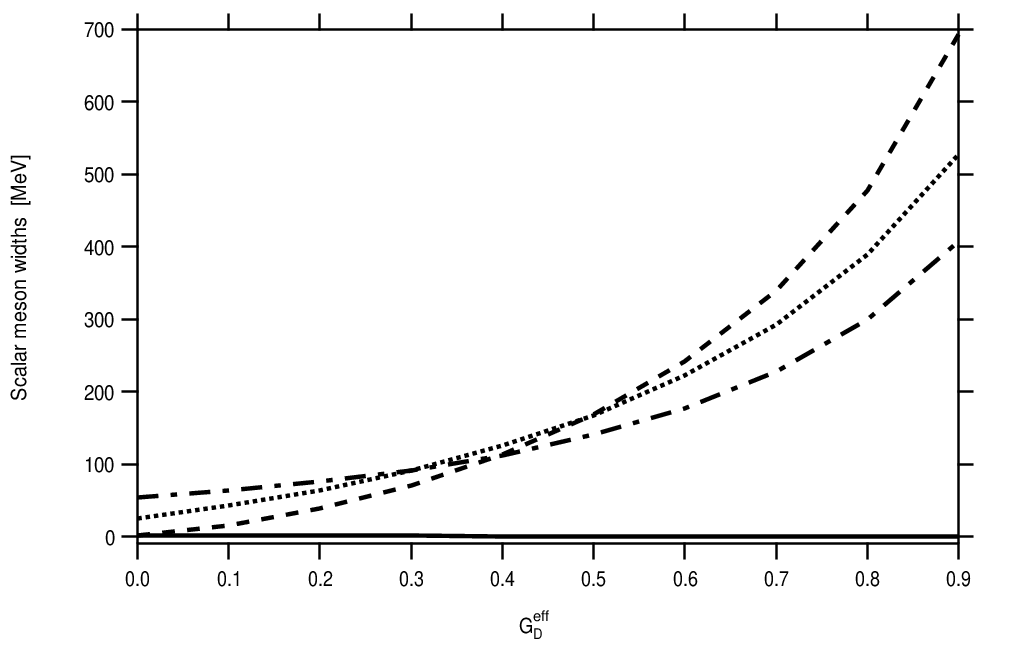}}
\caption{The calculated $q\qbar$ decay widths of the scalar mesons 
as functions of the effective coupling constant $G_D^{\rm eff}$ of 
the \UA\ breaking KMT interaction.
The solid, dashed, dotted and dash-dotted lines represent 
$m_\sigma$, $m_{a_0}$, $m_{K_0^*}$ and $m_{f_0}$, respectively.}
\label{fig:width}
\end{figure}

\begin{figure}[t]
	\centerline{\epsfxsize= 10cm \epsfbox{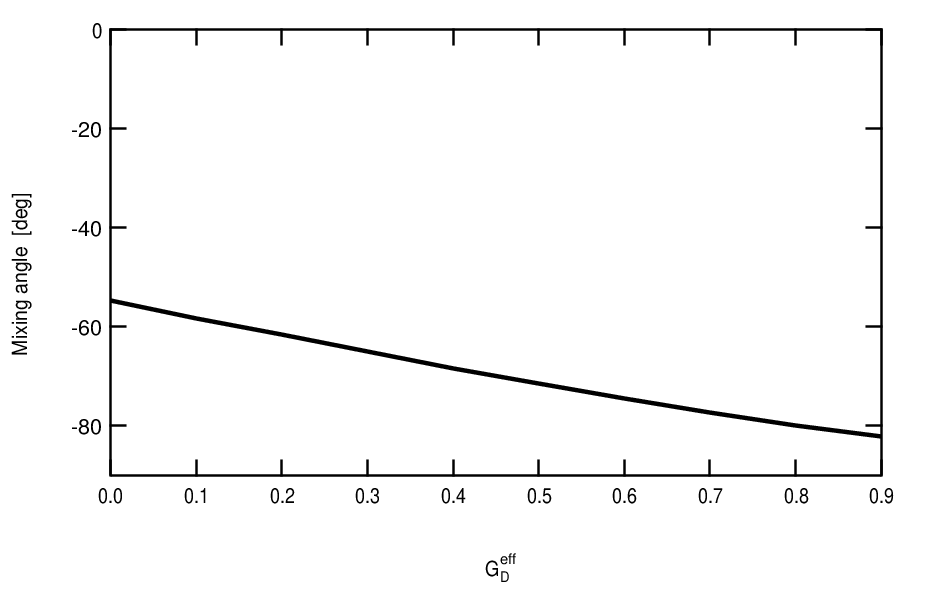}}
\caption{The calculated mixing angle of the $\sigma$ meson as 
a function of the effective coupling constant $G_D^{\rm eff}$ of 
the \UA\ breaking KMT interaction.}
\label{fig:mix}
\end{figure}

Let us now discuss our results of the scalar mesons. 
The calculated results of the scalar-meson masses, $q\qbar$ decay widths and 
the mixing angle $\theta$ are shown in Fig. \ref{fig:mass}, 
Fig. \ref{fig:width} and Fig. \ref{fig:mix}, respectively.
The $q\qbar$ decay widths of the scalar mesons shown there are 
unphysical ones.  We present them just for showing the 
pole positions in the complex $q^2$-plane.
When $G_{D}^{\rm eff}$ is zero, our lagrangian does not cause the flavor 
mixing and therefore the ideal mixing is achieved. 
The $\sigma$ is purely $u\bar u + d\bar d$, which corresponds to 
$\theta = -54.7^{\circ}$, and is degenerate to the $a_0$ in this limit.
When one increases the strength of the \UA\ breaking KMT interaction, 
the $q\qbar$ attraction in $\sigma$ increases and $\sigma$ state moves 
from the ideal mixing state toward the flavor singlet state. 
It means that the strange quark component of $\sigma$ increases as 
$G_D^{\rm eff}$ becomes bigger.  
Since the increase of the attractive force compensates with the increase 
of the strange quark component, 
$m_\sigma$ is almost independent of the strength of the \UA\ breaking 
interaction. 
The $\qbar q$ decay width of the $\sigma$ meson is very small, i.e., 
less than 2 MeV and therefore we neglect it in our calculation of 
the mixing angle.
At $G_{D}^{\rm eff} = 0.7$, the calculated mixing angle is
$\theta = -77.3^\circ$, corresponding to  about 15\% mixing of 
the strangeness component in $\sigma$.

Hatsuda and Kunihiro have discussed the masses and mixing angle of the 
isoscalar nonstrange ($\sigma_{NS}$) and strange ($\sigma_S$) scalar mesons 
using the similar model.\cite{HK1994} They have reported a rather small mixing 
between $\sigma_{NS}$ and $\sigma_S$. The reason of the difference between 
their result and our result is the strength of the $U_A(1)$ breaking KMT
interaction.  The strength of the $U_A(1)$ breaking KMT interaction used in 
the present study is much stronger than that used in their study. 
They have determined the strength from the $\eta'$ mass, while we have 
fixed it from the radiative decays of $\eta$. Strong $U_A(1)$ breaking
interaction suggests that the instanton liquid picture of the QCD vacuum.%
\cite{Instanton} 
In Ref.~\cite{HK1994}, they have discussed the origin of the difference of
the mixing properties of the scalar mesons and pseudoscalar mesons. 
We agree with their qualitative discussion, namely, the flavor mixing 
between $\sigma$ and $f_0$ is weaker than that between 
$\eta$ and $\eta'$.
Shakin has also pointed out that the KMT interaction mixes the 
$\sigma_{NS}$ and $\sigma_S$, while he assigned the lowest $I=0$ $q\qbar$ 
state to $f_0(980)$. \cite{Sha2002}

Let us turn to the discussion of the $a_0$ and $K_0^*$ mesons.
As shown in  Fig. \ref{fig:mass}, both $m_{a_0}$ and $m_{K_0^*}$ 
increase as $G_D^{\rm eff}$ increases. The slope for $m_{a_0}$ is 
steeper than that for $m_{K_0^*}$, 
which is consistent with the simple argument based 
on the form of the effective interaction Eq. (\ref{ekmt}).
At $G_{D}^{\rm eff} = 0.7$, the calculated masses are $m_{a_0} = 816$ MeV and
$m_{K_0^*} = 1002$ MeV, therefore
the \UA\ breaking interaction pushes up the $a_0$ and $K_0^*$ masses about 
161 MeV and 88 MeV, respectively. 
Although the effect of the \UA\ breaking interaction on the $K_0^*$ meson 
is smaller than that on the $a_0$ meson, our numerical results show that 
it is not enough to support the existence of the light $\kappa$ state.

As for the $f_0$ meson, we have shown our results in 
Figs. \ref{fig:mass} and \ref{fig:width}.  At $G_{D}^{\rm eff} = 0$, 
the $f_0$ state is expected to be pure $s\sbar$ state in our model.
Because of the $q\qbar$ decay width, we cannot calculate the 
mixing angle for $f_0$.  The calculated mass of the $f_0$ meson 
at $G_{D}^{\rm eff} = 0$ is $m_{f_0} = 1.163$ GeV which is above the 
$s\sbar$ threshold $2 M_s = 1.113$ GeV. As shown in Ref.~\cite{Dmitra}, 
the symmetry breaking effect by the current quark mass term pushes up the 
scalar meson mass above the $q\qbar$ threshold and the following relation 
is obtained by using the bosonization technique with the lowest order
derivative expansion in the NJL model.
\begin{equation}
\left( m_{\rm scalar \, meson}^2 - ({q\bar{q} \, {\rm threshold \,energy})}^2 
\right) 
\propto m_{\rm current \, quark}
\end{equation}
Our results at $G_{D}^{\rm eff} = 0$ are $m_{a_0}^2 - 4 M_u^2 = 0.008$ 
GeV$^2$, 
$m_{K_0^*}^2 - (M_u + M_s)^2 = 0.060$ GeV$^2$ and 
$m_{f_0}^2 - 4 M_s^2 = 0.115$ GeV$^2$, respectively.  The above 
simple mass relation therefore holds in our case too.
Fig. \ref{fig:mass} shows that the $f_0$ meson mass is almost independent
of the strength of the KMT interaction. The situation is just opposite to the 
$\sigma$ case, i.e., the increase of the repulsive force by the KMT interaction
compensates with the decrease of the strange quark component of the $f_0$ meson
when one increases the strength of the KMT interaction.

It should be noted here that 
in the $SU_L(3) \times SU_R(3)$ version of 
linear sigma model, not only the three-meson flavor 
determinant term but also the chiral invariant 
four-meson terms give rise to the $\sigma - a_0$ mass difference.
\cite{CH1974,BFMNS2001} 
We note that the extended NJL model does not give such type of interaction.

\chapter{Formulation  of the Improved Ladder Approximation Approach }

In this chapter, we present the formulation of 
the Improved  Ladder Approximation (ILA) of QCD. 
The approximation is applied to the Schwinger-Dyson (SD) 
equation for the quark propagator and to the 
Bethe-Salpeter (BS) equation for the pseudoscalar and 
scalar meson.

\section{Lagrangian}\label{formalism_model}
The ladder approximation is the lowest order in the perturbation 
theory for the interaction kernels employed in this approximation may not 
be valid except for a heavy quark systems, where the 
coupling constant is small for this approximation. 
It, however, can be improved at low mass region 
by including a certain set of  higher-order 
diagrams which bring 
the running coupling constant, $\alpha_S(p^2)$.

\begin{figure}[tbh]
  \centerline{ \epsfysize= 10cm \epsfbox{ 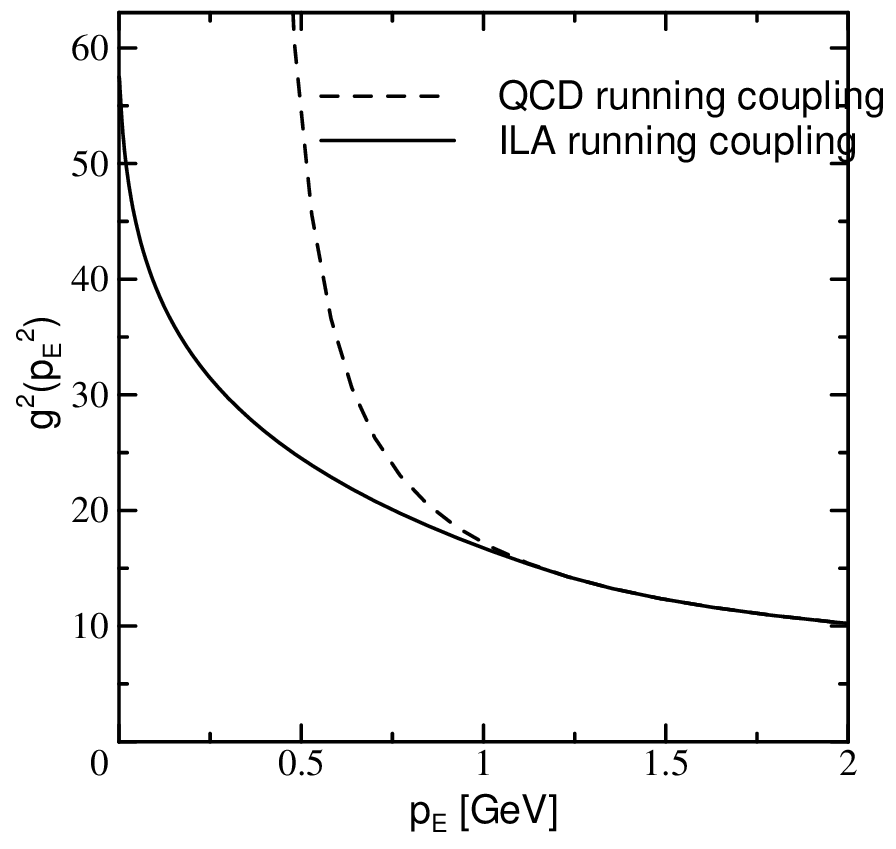} }
  \caption{The running coupling constant of the QCD and the ILA }
  \label{FIG:coupling}
\end{figure}

In the present study we employ the following Lagrangian 
density the ILA of QCD, 
\begin{eqnarray}
\label{Eq:totalL}
&&\L [\psi , \psib ] = \psib (i\dsla -m_0)\psi 
+\L _{\rm GE}[\psi,\psib ]+\L _{\rm KMT}[\psi,\psib ]\\
&&\psi = (u,d,s)^{T}. 
\end{eqnarray}
%
In (\ref{Eq:totalL}), 
$m_0$ denotes the bare  quark mass matrix $m_0 = \mbox{diag}(m_q,m_q,m_s) $ 
where we assume the isospin invariance. 
$\L _{\rm GE}$ denotes the gluon exchange interaction 
\begin{eqnarray}
\L _{\rm GE}[\psi,\psib ]
=
&&-\frac{1}{2} \int _{pp' qq'} \K ^{mm' ,nn'}(p,p',q,q')\nonumber \\
&&\qquad \qquad\times
\psib_{m}(p)\psi_{m'}(p')\psib_{n}(q)\psi_{n'}(q')
e^{-i(p+p'+q+q')x}
\end{eqnarray}
where 
\begin{eqnarray}
\K ^{mm' ,nn'}(p,p',q,q')
= i\bar{g}^2 D^{\mu \nu}\left(\frac{p+p'}{2}-\frac{q+q'}{2} \right)
(\gamma_{\mu}T^a)^{mm'}(\gamma_{\nu}T^a)^{nn'}.
\end{eqnarray}
To reduce the expressions we use an abbreviation 
$\int_p \def \int \frac{d^4p}{(2\pi)^4}$ through this 
chapter. 
The indeces $m,n,\cdots$ represent the components of the color, flavor and 
Dirac space. 
In the gluon propagator we employ the Landau gauge, 
\begin{eqnarray}
iD^{\mu \nu}(k) = \left( g^{\mu \nu} -\frac{k^{\mu}k^{\nu}}{k^2} \right)
\frac{-1}{k^2}
\end{eqnarray}
and the Higashijima-Miransky type running coupling constant $\bar{g}^2$ 
defined as follows.
\begin{eqnarray}
\bar{g}^2(p^2_E,q^2_E) 
= \theta(p^2_E-q^2_E)g^2(p^2_E)+ \theta(q^2_E-p^2_E)g^2(q^2_E)
\end{eqnarray}
with
\begin{eqnarray}
\label{runningCoupling}
&&g^2(p_E^2)=
\left\{
\begin{array}{ll}
\frac{1}{\beta_0}\frac{1}{1+t}  \mbox{for \quad } \tIF \le t \\
&\\
\frac{1}{2\beta_0}\frac{1}{(1+\tIF)^2}
\left[ 
      3 \tIF -t_0 +2 
      -\frac{(t-t_0)^2}{\tIF-t_0}
\right]
 & \mbox{for} \quad t_0 \le t \le \tIF \\
&\\
\frac{1}{2\beta_0}
\frac{3 \tIF -t_0 +2}{(1+\tIF)^2}
 & \mbox{for} \quad t \le t_0  \\
\end{array}
\right. \\
&&t=\ln \frac{p_E^2}{\lamQ^2} -1 \\
&&\beta_{0} = \frac{1}{(4 \pi)^2}\frac{11N_C-2N_f}{3}
\end{eqnarray}
Here, $p_E$ and $q_E$ denote the Euclidian momenta defined by 
\begin{eqnarray}
p = (p_0, \vec{p}) \quad &\to& \quad p_E = (ip_4, \vec{p}) \\
p^2 = p_0^2 - \vec{p}^2 \quad &\to& \quad p^2_E = -p^2 = \vec{p}^2 + p_o^2 
\end{eqnarray}
In eq.($\ref{runningCoupling}$) the infrared cut-off $\tIF$ 
is introduced. Above  $\tIF$ $g^2(p_E^2)$ develops according to 
the one loop solution of the QCD  renormalization group equation, 
while 
below $t_0$, $g^2(p_E^2)$ is kept constant. 
These two regions are connected by a quadratic polynomial so that 
$g^2(p_E^2)$ becomes a smooth function. Here $N_C$ and $N_f$ are 
the number of colors and active flavors respectively. 
We use $N_C=N_f=3$ in this study. 
The behavior of the running coupling constant is shown in the 
Fig. \ref{FIG:coupling} together with the one-loop QCD running coupling 
constant for $\lamQ = 600[\MeV]$. 

$\L_{\rm KMT}$ is the Kobayashi-Maskawa-'t Hooft (KMT) interaction 
given by 
\begin{eqnarray}
\L_{\rm KMT}= &&-\frac{1}{3}G_D\epsilon^{f_1f_2f_3}\epsilon^{g_1g_2g_3}
\bar{w}(\del_{x_1};\del_{y_1};\del_{x_2};\del_{y_2};\del_{x_3};\del_{y_3})
\nonumber \\
&&\times
\left\{
[\psib_{g_1}(x_1)\psi_{f_1}(y_1)][\psib_{g_2}(x_2)\psi_{f_2}(y_2)]
[\psib_{g_3}(x_3)\psi_{f_3}(y_3)]
\right.
\nonumber \\
&&+3
\left.
[\psib_{g_1}(x_1)\psi_{f_1}(y_1)][\psib_{g_2}(x_2)\gamma_5\psi_{f_2}(y_2)]
[\psib_{g_3}(x_3)\gamma_5\psi_{f_3}(y_3)]
\right\}
\Bigg{|}_{*}
\end{eqnarray}
where $f_1,g_1,\cdots$ are flavor indices, $\epsilon $ denotes 
the antisymmetric tensor with $\epsilon^{uds}=1$ and the 
symbol $*$ at the end of the equation means to take the limit 
$x_1,x_2,\cdots \to x$ 
after all derivatives are operated. 
This interaction causes the $U_A(1)$ symmetry breaking and also 
flavor mixing.

We introduce a weight function $\bar{w}$ which is necessary so 
that the KMT interaction is turned off at the high energy region. 
Then the asymptotic behavior of the ILA are kept. 
We use the 
following Gaussian form

\begin{eqnarray}
&&
\bar{w}(\del_{x_1};\del_{y_1};\del_{x_2};\del_{y_2};\del_{x_3};\del_{y_3})
=\bar{w}
\left( 
       \frac{\del^2_{x_1}+\del^2_{x_2}+\del^2_{x_3}
             +\del^2_{y_1}+\del^2_{y_2}+\del^2_{y_3}}
             {2} 
\right) \\
&&w(\mu^2) = \exp (-\kappa \mu^2)
\end{eqnarray}
This weight function is convenient for numerical calculations as it 
is factorized as 
\begin{eqnarray}
w (-p^2-q^2)  = w(-p^2)w(-q^2). 
\end{eqnarray}

The parameter $\kappa$ is taken as 
\begin{eqnarray}
\kappa  = 0.7 [\GeV^{-2}].
\end{eqnarray}
This value corresponds to the form factor of the instanton 
of the average size $\rho$, about $1/3$[fm].
The instanton form factor,
\begin{eqnarray}
\frac{1}{x_E^2+\rho^2} \propto 1-\frac{x^2_E}{\rho^2} + \cdots
\end{eqnarray}
can be identified for small $x_E^2$ 
with the Fourier transformation of the weight function
\begin{eqnarray}
\mbox{F.T.} w(q^2_E) = C \exp\left(-\frac{x^2_E}{4\kappa}\right)
\propto 1-\frac{x^2_E}{4\kappa} + \cdots
\end{eqnarray}
with 
\begin{eqnarray}
4\kappa = \rho^2.
\end{eqnarray}
\begin{figure}[tbh]
  \centerline{ \epsfysize= 4cm \epsfbox{ 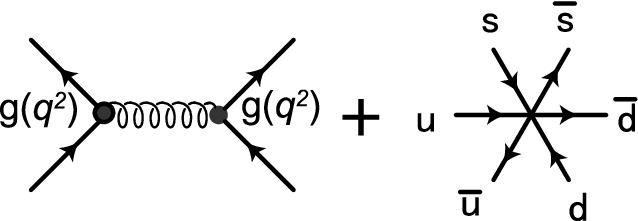} }
  \caption{The interactions of the ILA. }
  \label{FIG:ILA_interaction}
\end{figure}

Finally we introduce the ultraviolet 
momentum cutoff $\lamUV$, which 
is necessary to regularize loop integrals 
in the ILA approach. 
It is useful to introduce the cutoff function 
\begin{eqnarray}
\frac{1}{f(p^2)} =
\left\{
\begin{array}{cc}
1 & p^2 < \lamUV ^2\\ 
0 & p^2 > \lamUV ^2
\end{array}
\right.
\end{eqnarray}
By adding it in the kinetic term as 
\begin{eqnarray}
\psib f(\del^2)(i\dsla -m_0)\psi
\end{eqnarray}
all integrals are regularized. 
We choose a very large $\lamUV$ ($\sim 100[\GeV]$) 
so that the results are not sensitive 
to the value of $\lamUV$. 

It should be noted that because the cut off function, 
f and w, depend on the momentum, 
the Noether currents are modified 
from the bare QCD currents.

\section{Renormalization of Quark Masses}

In our approach, the parameters of the Lagrangian are 
renormalized  at the cutoff scale 
$\lamUV$. 
Since the $\lamUV$ should be taken 
enough large, it may not be  appropriate 
to fix the quark mass at such high energy.  
Instead of the quark mass ${m_q}_0$, ${m_s}_0$ 
at the scale $\lamUV$, 
we choose renormalized the quark mass at the momentum scale 
$\mu = 2 [\GeV]$ as ${m_q}_R = 5 [\MeV]$ and 
${m_s}_R = 100 [\MeV]$ and calculate the ${m_q}_0$ and ${m_s}_0$ 
in the following way. 

The renormalization constants $Z_{m_q}$ and $Z_{m_s}$  are defined by 
\begin{eqnarray}
m_q = Z_{m_q}^{-1}{m_q}_R , \quad 
m_s = Z_{m_s}^{-1}{m_s}_R.
\end{eqnarray}
We take the renormalization condition as 
\begin{eqnarray}
&&\frac{\del B_q(\mu^2)}{\del {m_q}_R}\Bigg|_{m_{qR}}=0\\
&&\frac{\del B_s(\mu^2)}{\del {m_s}_R}\Bigg|_{m_{sR}}=0.
\end{eqnarray}

We define the  quark condensate as follows 
\begin{eqnarray}
&&<:\psib(0) \psi(0):> = -\int _p \Tr[S_F^{\rm Full}(p)] 
+ \int_p Tr[S_F^{\rm bare}]
\end{eqnarray}
where
\begin{eqnarray}
&&\int _p \Tr[S_F^{\rm Full}(p)] = N_C\int_p \frac{4iB(p^2)}{p^2-B^2(p^2)}
\\
&&\int _p \Tr[S_F^{\rm bare}(p)]
 = N_C\int_p \frac{4i\frac{\del B}{\del m_R}m_R}
 {p^2-\left(\frac{\del B}{\del m_R}m_R \right)^2}
\end{eqnarray}
%

\section{Effective Action}\label{EffectiveAction}
To derive the Schwinger-Dyson (SD) equation and the Bethe-Salpeter (BS)
 equation, 
we use the Cornwall-Jackiw-Tomboulis (CJT) effective action 
formulation \cite{CJT1974}. 

The CJT effective action is defined as 
\begin{eqnarray}
\label{CJTaction}
\Gamma[S_F] = i \Tr \mbox{Ln}[S_F] - i \Tr[S_0^{-1} S_F] 
+ \Gamma_{\rm loop}[S_F]
\end{eqnarray}
The last term of Eq.$(\ref{CJTaction})$ is the residual term, 
and is 
given by the sum of all the Feynman amplitudes of corresponding to the 
two particle irreducible vacuum diagrams 
with two ore more loops. 
In these amplitudes, 
quark 
propagator 
is substituted  by the full propagator
\begin{eqnarray}
S_F(x,y) = \langle 0 | T\psi(x)\psib(y)| 0\rangle.
\end{eqnarray}
In this study, we employ the lowest loop order 
and the leading $1/N_C$ contribution for 
the $\Gamma_{\rm loop}[S_F]$. 
Then the  CJT action becomes 
\begin{eqnarray}
\Gamma[S_F] = i \Tr \mbox{Ln}[S_F] - i \Tr[S_0^{-1} S_F] 
+ \Gamma_{\rm GE}[S_F] + \Gamma_{\rm KMT}[S_F]
\end{eqnarray}
where 
\begin{eqnarray}
\Gamma_{\rm GE}[S_F] = 
&& -\frac{1}{2} \int d^4x 
{\cal K}^{m_1m_2,n_1n_2}(i\del_{x_1},i\del_{x_2};i\del_{y_1},i\del_{y_2})
\nonumber \\
&& \times \left(
    {S_F}_{m_2m_1}(x_2,x_1){S_F}_{n_2n_1}(y_2,y_1)
   -{S_F}_{m_2n_1}(x_2,y_1){S_F}_{n_2m_1}(y_2,x_1)
   \right)
   \Bigg|_{*}
\end{eqnarray}
\begin{eqnarray}
\Gamma_{\rm KMT}[S_F] 
=&& -\frac{G_D}{3}\int d^4x \epsilon^{f_1f_2f_3g_1g_2g_3}
w\left(\frac{\del^2_{x_1}+\del^2_{y_1}+\del^2_{x_2}+\del^2_{y_2}+\del^2_{x_3}+\del^2_{y_3}}{2}\right)
\nonumber \\
&& \times \left( -\tr^{\rm (DC)}[{S_F}_{f_1g_1}(y_1,x_1)]
                  \tr^{\rm (DC)}[{S_F}_{f_2g_2}(y_2,x_2)]
                  \tr^{\rm (DC)}[{S_F}_{f_3g_3}(y_3,x_3)]
\right.
\nonumber \\
&& \qquad  \left.-3\tr^{\rm (DC)}[{S_F}_{f_1g_1}(y_1,x_1)]
                  \tr^{\rm (DC)}[\g5{S_F}_{f_2g_2}(y_2,x_2)]
                  \tr^{\rm (DC)}[\g5{S_F}_{f_3g_3}(y_3,x_3)]
\right)\Bigg|_{*}.
\nonumber \\
\end{eqnarray}

It should be noted that 
 the global $SU_{L}(3) \times SU_{R}(3)$ 
symmetry is preserved 
within this approximation if the quark masses $m_q$ and 
$m_s$ vanish. 

In fact, the total effective action
$\Gamma[S_F] $ is invariant under the infinitesimal  
global chiral transformation 
\begin{eqnarray}
&&S_F(x,y) \to 
   \left(1+i\g5\frac{\lambda^a}{2}\theta^a\right)
   S_F(x,y)
   \left(1+i\g5\frac{\lambda^a}{2}\theta^a\right),
\end{eqnarray}
except for the quark mass term.

From this truncated effective action, the rainbow SD equation and 
the ladder BS equation are derived  consistently.

\section{Schwinger Dyson equation}

The Schwinger-Dyson equation is derived by the stability condition of 
the CJT action
\begin{eqnarray}
\frac{\delta \Gamma[S_F]}{\delta S_F(x,y)} = 0.
\end{eqnarray}

\begin{figure}[t]
  \centerline{ \epsfysize= 2.5cm \epsfbox{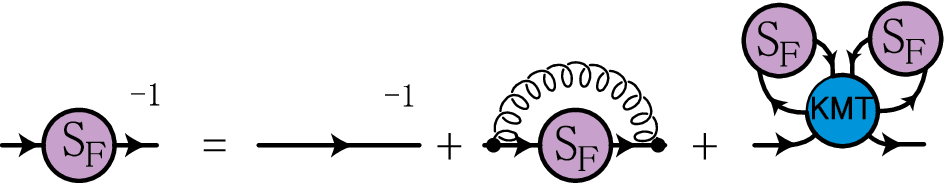} }
  \caption{The SD equation}
  \label{FIG_SDeq}
\end{figure}

Introducing the regularized propagators 
\begin{eqnarray}
&&S_F^R(q) = f(q^2)S_F(q)\\
&&S_0^R(q) = f(q^2)S_0(q),
\end{eqnarray}
the SD equation in the momentum space becomes 
\begin{eqnarray}
i{S_F^R}^{-1}(q) - i {S_0^R}^{-1} (q) &=&
- \frac{C_F}{f(-q^2)}\int_p \frac{1}{f(-p^2)}
\bar{g}^2(-q^2,-p^2)iD^{\mu\nu}(p-q)
\gamma_{\mu}S_F(p)\gamma_{\nu}
\nonumber\\
&&-G_D\Vec{1}_C\Vec{1}_f\epsilon^{gf_1f_2}\epsilon^{fg_1g_2}
\int_{p,k}\frac{1}{f(-p^2)f(-k^2)}w(-q^2-p^2-k^2)
\nonumber\\
&&\qquad\qquad\qquad
  \times \tr^{\rm (DC)}[{S_F^R(p)}_{g_1f_1}]
         \tr^{\rm (DC)}[{S_F^R(k)}_{g_2f_2}]
\end{eqnarray}
with the coefficient from  the color space,   
\begin{eqnarray}
C_F = \frac{\tr[T^aT^a]}{N_C} = \frac{N_C^2 -1}{2N_C}. 
\end{eqnarray}
This equation is shown diagrammatically in Fig.$(\ref{FIG_SDeq})$. 

Generally the quark propagator is parametrized as 
\begin{eqnarray}
S_F^R(q) = \frac{i}{\qsla A(q^2)-B(q^2)}. 
\end{eqnarray}
In the Landau gauge, it can be shown that the solution satisfies 
$A(-q_E^2)=1$. 
Then the SD equation becomes the integral equation only of the
mass function $B(q^2)$ and reads 
\begin{eqnarray}
B_q(-q^2) = &&m_q 
+\frac{3C_F}{16\pi^2}\int^{\lamUV^2}_0 dk^2 \bar{g}^2(q^2,k^2)
\left\{\theta(k^2-q^2)+\frac{k^2}{q^2}\theta(q^2-k^2)\right\}
\frac{B_q(-k^2)}{k^2 + B_q(-k^2)}
\nonumber \\
&&
+\frac{G_DN_C^2w(q^2)}{8\pi^4}
\int^{\lamUV^2}_0 dk^2 w(k^2)\frac{k^2B_q(-k^2)}{k^2 + B_q(-k^2)}
\int^{\lamUV^2}_0 dl^2 w(l^2)\frac{l^2B_s(-l^2)}{l^2 + B_s(-l^2)}
\nonumber \\
\\
B_s(-q^2) = &&m_s 
+\frac{3C_F}{16\pi^2}\int^{\lamUV^2}_0 dk^2 \bar{g}^2(q^2,k^2)
\left\{\theta(k^2-q^2)+\frac{k^2}{q^2}\theta(q^2-k^2)\right\}
\frac{B_s(-k^2)}{k^2 + B_s(-k^2)}
\nonumber \\
&&
+\frac{G_DN_C^2w(q^2)}{8\pi^4}
\int^{\lamUV^2}_0 dk^2 w(k^2)\frac{k^2B_q(-k^2)}{k^2 + B_q(-k^2)}
\int^{\lamUV^2}_0 dl^2 w(l^2)\frac{l^2B_q(-l^2)}{l^2 + B_q(-l^2)}.
\nonumber\\
\end{eqnarray}
%

\section{Bethe-Salpeter equation}
To treat the pseudoscalar mesons and the scalar meson as quark-antiquark 
bound state, we use the homogeneous Bethe-Salpeter equation. 
It is derived by 
\begin{eqnarray}
\frac{\delta^2 \Gamma[S_F]}{\delta {S_F}_{mn}(x,y)\delta {S_F}_{m'n'}(y',x')}
\chi_{n',m'}(y',x';P_B) = 0
\end{eqnarray}
where
\begin{eqnarray}
\chi_{n',m'}(y',x';P_B) 
= \langle 0|T \psi_{n'}(y')\psib_{m'}(x')|\Vec{P}\rangle
\end{eqnarray}
denotes the BS amplitude. The normalization condition is taken as 
\begin{eqnarray}
\langle \Vec{P}_B | \Vec{P'_B} \rangle
= (2\pi)^{3} 2{P_B}_0 \delta^{3}(\Vec{P}_B-\Vec{P}'_B)
\end{eqnarray}
and $P_B=\sqrt{M_B^2 + \Vec{P}_B^2}$ where $\Vec{P}_B$
 is the on-shell momentum of 
 the meson. 

\begin{figure}[t]
  \centerline{ \epsfysize= 2.5cm \epsfbox{ 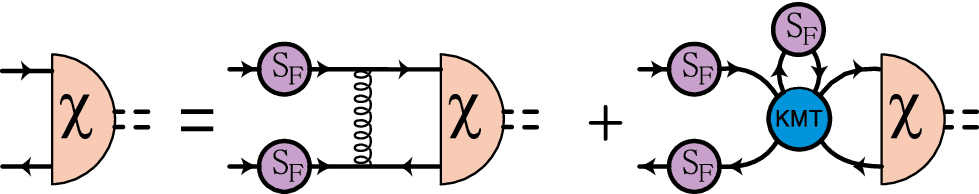} }
  \caption{The BS equation }
  \label{FIG_BSeq}
\end{figure}

Introducing the regularized BS amplitude by 
\begin{eqnarray}
&&
\chi^R_{n',m'}(q;P_B) 
 = f(-q_+^2)\chi_{n',m'}(q;P_B)  f(-q_-^2)\\
&&q_+ = q +\frac{P_B}{2}, \qquad q_- = q -\frac{P_B}{2}
\end{eqnarray}
the BS equation in momentum space becomes 
\begin{eqnarray}
&&S_F^{R^ {-1}}(q_+)\chi^R(q;P)S_F^{R^{ -1}}(q_-)\nonumber\\
&&=-iC_F \int_k \frac{1}{f(-k_+^2)f(-k_-^2)}\bar{g}^2(-q^2,-k^2) 
i D^{\mu \nu}(q-k)\gamma_{\mu}\chi^R(k;P)\gamma_{\nu}\nonumber\\
&& - 2iG_D\epsilon^{ghf'}\epsilon^{fh'g'}\Vec{1}_{C}
\int_{p,k}\frac{1}{f(-p^2)f(-k_+^2)f(-k_{-}^2)}
w \left( -p^2 - q^2 - k^2 -\frac{P_B^2}{2}  \right)
\nonumber \\
&&\times \tr^{(DC)}[{S_F^R}_{h'h}(p)]
\left\{
  (\g5)\tr^{(DC)}[\g5 \chi^R_{g'f'}(k;P_B)] 
  + (1) \tr^{(DC)}[\chi^R_{g'f'}(k;P_B)]
\right\}
\end{eqnarray}
which is shown diagrammatically  in Fig.($\ref{FIG_BSeq}$).

For the pion, the BS amplitude can be written in terms of 
four scalar amplitudes as in Ref.\cite{NNTYO2000},
\begin{eqnarray}
\chi^R_{nm}(k;P) =  \Vec{1}_{C} \frac{\lambda^{a}}{2}
\left( \phi_S(k;P) + \phi_P(k;P) \ksla + \phi_Q(k;P)\Psla
+\phi_T(k;P) \frac{1}{2}(\Psla \ksla - \ksla \Psla)\right)\g5
\end{eqnarray}
where $\lambda^a$ denotes the flavor structure of the pion state. 
The neutral pion, for instance,  is given by $a=3$
\begin{eqnarray}
\lambda^3 = 
\left(
\begin{array}{ccc}
1 &0 &0 \\
0 &-1 &0 \\
0 &0 &0 \\
\end{array}
\right)
\end{eqnarray}
On the other hand, for the $\eta$ and $\eta'$ mesons, the BS amplitudes are 
written in terms of eight scalar amplitudes,
\begin{eqnarray}
\chi^R_{nm}
=&&
   \Vec{1}_{C} \frac{\lambda^{q}}{2}
   \left( \phi^q_S(k;P) + \phi^q_P(k;P) \ksla + \phi^q_Q(k;P)\Psla
  +\phi^q_T(k;P) \frac{1}{2}(\Psla \ksla - \ksla \Psla)\right) \g5 
\nonumber \\
&+&
   \Vec{1}_{C} \frac{\lambda^{s}}{2}
   \left( \phi^s_S(k;P) + \phi^s_P(k;P) \ksla + \phi^s_Q(k;P)\Psla
  +\phi^s_T(k;P) \frac{1}{2}(\Psla \ksla - \ksla \Psla)\right)\g5 
\end{eqnarray}
where the flavor matrices $\lambda^{q}$ and $\lambda^{s}$ are defined by
\begin{eqnarray}
\lambda^q = 
\left(
\begin{array}{ccc}
1 &0 &0 \\
0 &1 &0 \\
0 &0 &0 \\
\end{array}
\right)
,\qquad
\lambda^s = 
\left(
\begin{array}{ccc}
0 &0 &0 \\
0 &0 &0 \\
0 &0 &\sqrt{2} \\
\end{array}
\right).
\end{eqnarray}
Because of the flavor mixing effects caused by the KMT interaction, 
the q and s components of the BS amplitude are mixed. 
We identify the ground state solution 
with the $\eta$ meson state and the 
first excited state with the 
$\eta'$ meson state. The explicit form of the 
BS equation is given  in the appendix.

Similarly the pseudoscalar mesons, the BS amplitude 
of the scalar meson is parametrized 
as 
\begin{eqnarray}
\chi^R_{nm}(k;P) =  \Vec{1}_{C} \frac{\lambda^{a}}{2}
\left( \phi_S(k;P) + \phi_P(k;P) \ksla + \phi_Q(k;P)\Psla
+\phi_T(k;P) \frac{1}{2}(\Psla \ksla - \ksla \Psla)\right)
\end{eqnarray}
for $a_0$ and 
\begin{eqnarray}
\chi^R_{nm}
=&&
   \Vec{1}_{C} \frac{\lambda^{q}}{2}
   \left( \phi^q_S(k;P) + \phi^q_P(k;P) \ksla + \phi^q_Q(k;P)\Psla
  +\phi^q_T(k;P) \frac{1}{2}(\Psla \ksla - \ksla \Psla)\right)
\nonumber \\
&+&
   \Vec{1}_{C} \frac{\lambda^{s}}{2}
   \left( \phi^s_S(k;P) + \phi^s_P(k;P) \ksla + \phi^s_Q(k;P)\Psla
  +\phi^s_T(k;P) \frac{1}{2}(\Psla \ksla - \ksla \Psla)\right)
\end{eqnarray}
for $\sigma$ or $f_0$, respectively.

In the numerical computation, we treat the BS equation in the 
Euclidean momentum region.
Then the physical solution, which corresponds to negative $P_E^2$, 
is obtained only by extrapolation from the $P_E^2 > 0$ region. 
It can be done in the following way. 
First, we rewrite the Euclidean BS equation in the form 
\begin{eqnarray}
\label{directBS}
\phi_A(q;P_E) = \int _{k_E} M_{AB}(q_E;k_E;P_E) \phi_B(k;P_E)
\end{eqnarray}
where $\phi_A$ and $\phi_B$ denotes the set of amplitude. 
This equation should not have solution at $P_E^2 > 0$ 
because, if it exists, it is a tachyon solution. 
Therefore we instead solve the eigenvalue equation 
\begin{eqnarray}
\lambda(P_E^2) \phi_A(q_E;P_E) = \int _k M_{AB}(q_E;k_E;P_E) \phi_B(k_E;P_E)
\end{eqnarray}
for a fixed $P_E^2 > 0$. 
Then we extrapolate the eigenvalue to $P_E^2 < 0$ 
as a function of $P^2_E$ 
and search for the on-shell point $\lambda(-M_B^2) = 1$. 
We employ the quadratic function of $P_E^2$ 
for the extrapolation. 
We also check the  results using the PCAC relation.

\section{Normalization and Decay Constants}

\subsection{Normalization}

In order to obtain the decay constants of the pseudoscalar mesons, 
it is necessary  to normalize the 
BS amplitude derived from the inhomogeneous BS 
equation. 
The normalization condition of the BS amplitude is given by 
\begin{eqnarray}
i \int _q \frac{1}{f(-q_+^2)f(-q_-^2)}
\chi^R_{n_1m_1}(q;P)\bar{\chi}^R_{m_2n_2}(q;P)
P^{\mu}\frac{\del}{\del P^{\mu}}
\left\{{S_F^R}_{n_2n_1}^{-1}(q_+){S_F^R}_{m_1m_2}^{-1}(q_-) \right\}
=-2P^2.
\end{eqnarray}
The explicit form of  this equation is shown in the appendix.

\subsection{Calculation of the Decay Constant}

Using the normalized BS amplitude, the decay constants 
of the pseudoscalar mesons 
are  obtained by 

\begin{eqnarray}
f_B = \frac{1}{P_B^2}\int_q && \frac{1}{f(-q_-^2)f(-q_+^2)}
\nonumber \\
&&\times
\Tr
  \left[
    \bar{\chi}^{R}(q;P_B)
      \left\{ i\g5\frac{\lambda^{\alpha}}{2}
             \left(
               \frac{f(-q_-^2)+f(-q_+^2)}{2}\Psla 
               +(f(-q_+^2)-f(-q_-^2))\qsla
             \right)
\right. \right. 
\nonumber \\
&&
\left. \left. 
 \qquad \quad
             +E^{\alpha}(q;P)
             +F^{\alpha}(q;P)
      \right\}
   \right]
\nonumber \\
\end{eqnarray}
\begin{eqnarray}
E^{\alpha}_{mn}(q;P)
=\int_k && \left[\left\{ K^{n'nmm'}_{(-k,q-\frac{P}{2},-q-\frac{P}{2},k+P)}
                        -K^{n'nmm'}_{(-k,q-\frac{P}{2},-q+\frac{P}{2},k)}
                 \right\}
                 \left(i\g5 \flavor S_F(k)\right)_{m'n'}
           \right.
\nonumber \\
           && +
           \left.
                 \left\{ K^{n'nmm'}_{(-k+P,q-\frac{P}{2},-q-\frac{P}{2},k)}
                        -K^{n'nmm'}_{(-k,q+\frac{P}{2},-q-\frac{P}{2},k)}
                 \right\}
                 \left(S_F(k) i\g5 \flavor \right)_{m'n'}
           \right]
\nonumber \\
\end{eqnarray}
\begin{eqnarray}
F^{\alpha}_{mn}(q;P) &=& - 2G_D \g5 \Vec{1}_{color} \int _{k,l}
\frac{1}{f(-k^2)f(-l^2)}\tr^{(DC)}[{S_F^R}_{g_2f_2}(k)]
\nonumber \\
&&\quad
  \times
  \left[
    \left\{
       w\left(-(q-\frac{P}{2})^2 - k^2 - l^2 \right)
      -w\left(-q^2 -\frac{3}{4}P^2 - k^2 - l^2 +Pl\right)
    \right\}
  \right.
\nonumber \\
&&\quad \quad
  \times \epsilon^{ff_2f_3}\epsilon^{gg_2g_3}
         \tr^{(DC)}\left[{S_F^R}(l) i\frac{\lambda^{\alpha}}{2}\right]_{g_3f_3}
\nonumber \\
&&\quad \quad+
    \left\{
       w\left(-(q+\frac{P}{2})^2 - k^2 - l^2 \right)
      -w\left(-q^2 -\frac{3}{4}P^2 - k^2 - l^2 -Pl\right)
    \right\}
\nonumber \\
&&\quad \quad
  \left.
  \times \epsilon^{ff_2f_3}\epsilon^{gg_2g_3}
         \tr^{(DC)}\left[ i\frac{\lambda^{\alpha}}{2} {S_F^R}(l)\right]_{g_3f_3}
  \right]
\end{eqnarray}
The terms $E^{\alpha}(q;P)$ and $F^{\alpha}(q;P)$ 
represent the corrections to the Noether axialvector current 
due to the momentum dependencies interaction. 

For the $\eta$ and $\eta'$ mesons, the flavor structure of the 
BS amplitude depends on the relative and total momenta in general. 
Therefore we can not fix $\lambda^{\alpha}$ in the 
definition  of the decay constant from the 
flavor structure of the BS amplitude. 
Instead we only consider the decay constants associated with the 
octet and singlet axialvector currents for the 
$\eta$, $\eta'$ mesons, i.e., 
we define and calculate 
$f_8^{\eta}$, $f_0^{\eta}$
,$f_8^{\eta'}$ and $f_0^{\eta'}$.

Since the BS equation is homogeneous, the overall sign of the BS amplitude, 
and therefore the decay constants, cannot be determined. We choose the sign 
as 
\begin{eqnarray}
\left\{
\begin{array}{l}
f^8_{\eta} > 0 \qquad \mbox{for} \qquad \eta \\
f^0_{\eta'} > 0 \qquad \mbox{for} \qquad \eta' \quad.\\
\end{array}
\right.
\nonumber
\end{eqnarray}
%

\section{PCAC relation}

In our approach, there are eight axial-vector currents, 
$J_{5\mu}^{\alpha}(\alpha = 1,\cdots,8)$ 
, which satisfy the PCAC relation

\begin{eqnarray}
\label{currentPCAC}
&&\del^{\mu}J_{5\mu}^{\alpha}(x) = 2\left[m_0J_5\right]^{\alpha}(x)
\\
&&\left[m_0J_5\right]^{\alpha}(x) = 
   \psib i\g5 
   \frac{f(\stackrel{\leftarrow}{\del^2}) m_0\lambda^{\alpha} 
           + \lambda^{\alpha}m_0 f(\del^2)}{4}
   \psi.
\end{eqnarray}
%

We take the matrix element of both the sides of eq. ($\ref{currentPCAC}$) 
between 
a meson state $\langle \Vec{P}_B |$ and the vacuum $| 0 \rangle$ 
and obtain 
\begin{eqnarray}
-f_B^{\alpha} M_B^2 = 2[m_0{\cal E}_B^{\alpha}] 
\end{eqnarray}
where $f_{B}^{\alpha}$ means the decay constant 
$f_{\pi},f_8^{\eta},f_8^{\eta'}$. 
$[m_0{\cal E}_B^{\alpha}]$
is defined by 

\begin{eqnarray}
[m_0{\cal E}_B^{\alpha}]
= \lim_{P \to P_B}
i\int _q \frac{1}{2}\left(\frac{1}{f(-q_-^2)}+\frac{1}{f(-q_+^2)}\right)
\tr\left[\bar{\chi}^R(q;P)m_0\g5\frac{\lambda^{\alpha}}{2}\right]
\end{eqnarray}
%

%

%

Since  this relation must be kept on the mass shell,  
we can use this as a check of the obtained mass 
by using the extrapolation. 
We introduce the ratio R defined by 
\begin{eqnarray}
\label{Eq:RValue}
R^{\alpha}(P_E^2) = 
\frac{f^{\alpha}_B(P_E^2)P_E^2}{2[m_0{\cal E}_B]^{\alpha}(P_E^2)
          }. 
\end{eqnarray}
As this ratio should become unity at $P_E^2 = -M_B^2$,  
the value of $R^{\alpha}(-M_E^2)$ is a good 
measure of the accuracy of the calculation, 
especially the extrapolation from the 
$P_E^2 > 0$ region to $P_E^2 = -M_B^2$ point.

\chapter{Results and Discussion}
In this chapter, we show the numerical results 
and give the disscussion. 

\section{Parameters}
%
In the present approach, there are eight input parameters:
 quark mass 
${m_q}_0$ for the up and down quarks,
 the current quark mass 
${m_s}_0$ for the strange quark,
 the scale parameter of QCD 
running coupling constant  $\lamQ$, the 
infrared cutoff $\tIF$, the smoothness parameter $t_0$, 
the strength parameter of the KMT $G_D$, the parameter 
of the weight function of the KMT $\kappa$ 
and the ultraviolet cutoff $\lamUV$.
We take the ultraviolet cutoff $\lamUV = 100[\GeV]$ and at such large value  
the physical observables do not depend on it. 
The quark masses ${m_q}_0, {m_s}_0$ are chosen 
so that the renormalized masses at the momentum scale $\mu = 2 [\GeV]$ 
become the ${m_q}_R = 4.5[\MeV]$ and ${m_s}_R = 150[\MeV]$, respectively. 
The $\kappa$ parameter is taken as $\kappa = 0.7 [\GeV^{-2}]$, 
which corresponds to the instanton of the average size $1/3[{\rm fm}]$ 
as shown in Sec.[$\ref{formalism_model} $]. 
The other parameters $\lamQ$, $\tIF$, $t_0$, $G_D$
are chosen as the pseudoscalar meson masses $M_{\pi}$, $M_{\eta}$, 
$M_{\eta'}$ and the pion decay constant $f_{\pi}$ 
are fitted to the experimental values. 
The parameters we used are $\lamQ = 600[\MeV]$, $t_0 = -6.89$, 
$\tIF = 0.204$ and $G_D = 75 [\GeV^{-5}]$. 

The results of the calculation with these parameters are 
$M_{\pi} = 136[\MeV]$, $M_{\eta} = 515[\MeV]$, $M_{\eta'} = 982 [\MeV]$ and 
$f_{\pi} = 95 [\MeV]$. 
These are in agreement with experimental values in less than $6\% $
 of deviation.
Table $\ref{TBL:Parameters}$ summarizes all the 
values of the parameters. 

Although $\lamQ$ is somewhat larger than the standard value 
$\lamQ = 100 \sim 300 [\MeV] $, a large $\lamQ$ 
is necessary in the ILA to generate 
the desired  dynamical chiral symmetry breaking (D$\chi$SB). 
It may indicate that  the chiral symmetry breaking 
is not fully explained by the 
ladder approximation of the 
one gluon exchange with $U_A(1)$ breaking effect. 
If we take a strong  KMT interaction, 
we may reduce $\lamQ$, 
but then $\eta - \eta'$ mass splitting becomes too large. 
As a result, we should choose a large $\lamQ$ so that 
the D$\chi$SB is generated mainly 
by the gluon exchange interaction. 
%
%
Although the results of our calculation show that the 
contribution of the KMT interaction to the quark mass function 
and to the quark condensate is small ($\le 10\%$), 
it gives  significant  effects on the mass spectrum and the mixing angle. 
Then this KMT interaction can be regarded to induce 
strong $U_A(1)$ breaking.

\begin{table}[btp]
\begin{center}
\begin{tabular}{|c|c|}  \hline
${m_q}_R(2[\GeV])$ &$4.5 [\MeV]$ \\
${m_s}_R(2[\GeV])$ & $150 [\MeV]$ \\
$\lamQ$ & $600 [\MeV]$\\
$\tIF$ & $ 0.204$\\
$t_0$ & $-6.89$\\
$G_D$ & $75 [\GeV^{-5}]$\\
$\kappa$ & $0.7 [\GeV^{-2}]$\\
$\lamUV$ & $100 [\GeV]$\\
\hline
\end{tabular} 
$\qquad \qquad$
\begin{tabular}{|c|c|}  \hline
$M_{\pi}$ & $136 [\MeV]$\\
$M_{\eta}$ & $515 [\MeV]$\\
$M_{\eta'}$ & $982 [\MeV]$\\
$f_{\pi}$ & $ 95 [\MeV]$\\
\hline
\end{tabular} 

\end{center}  
 \caption{The values of the the parameters in the ILA 
 approach and the obtained observable quantities.  
 }
 \label{TBL:Parameters}
\end{table}    
%

\section{Solution of the SD equation}

\begin{figure}[tbh]
  \centerline{ \epsfysize= 6cm \epsfbox {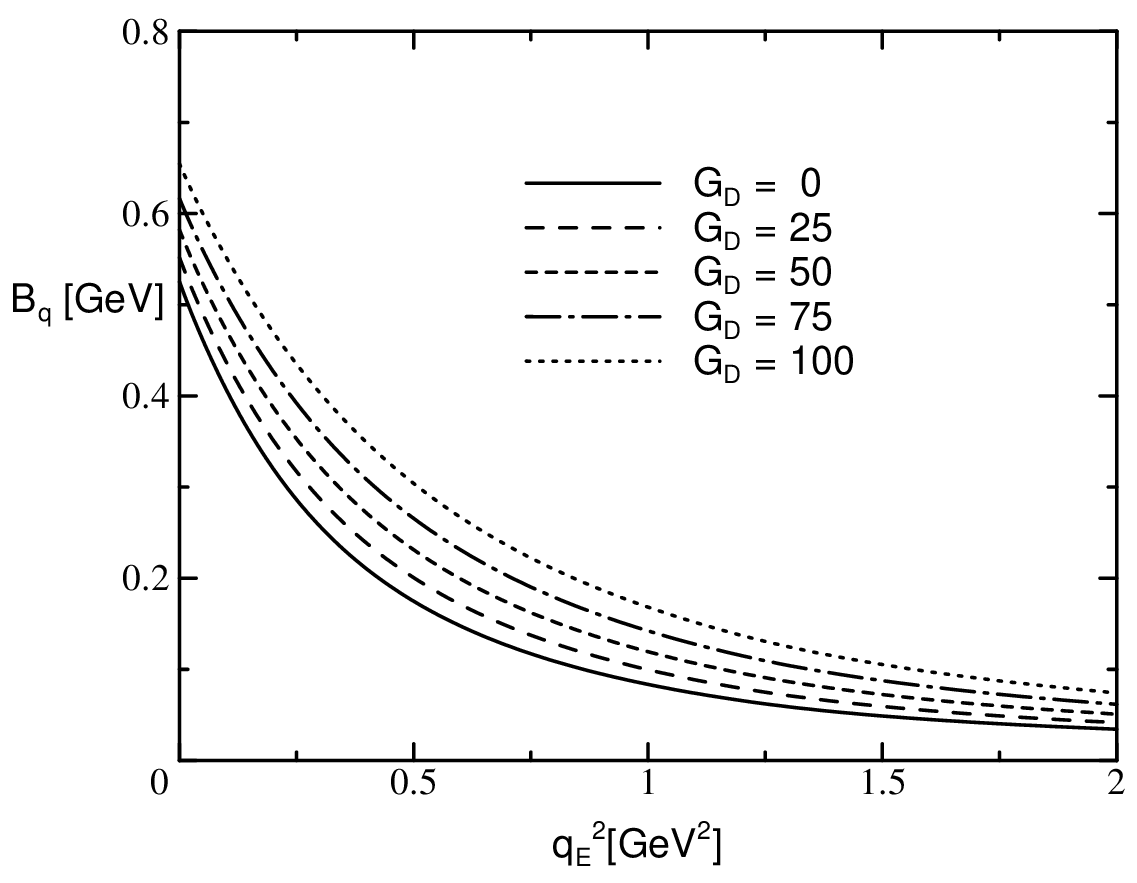} }
  \caption{$q_E^2$ dependences of the mass function of the q quark $B_q$ with 
           $G_D=0, 25, 50, 75,$ $100 [\mbox{GeV}^{-5}]$}
  \label{FIG:Bq}
\end{figure}

\begin{figure}[tbh]
  \centerline{ \epsfysize= 6cm \epsfbox {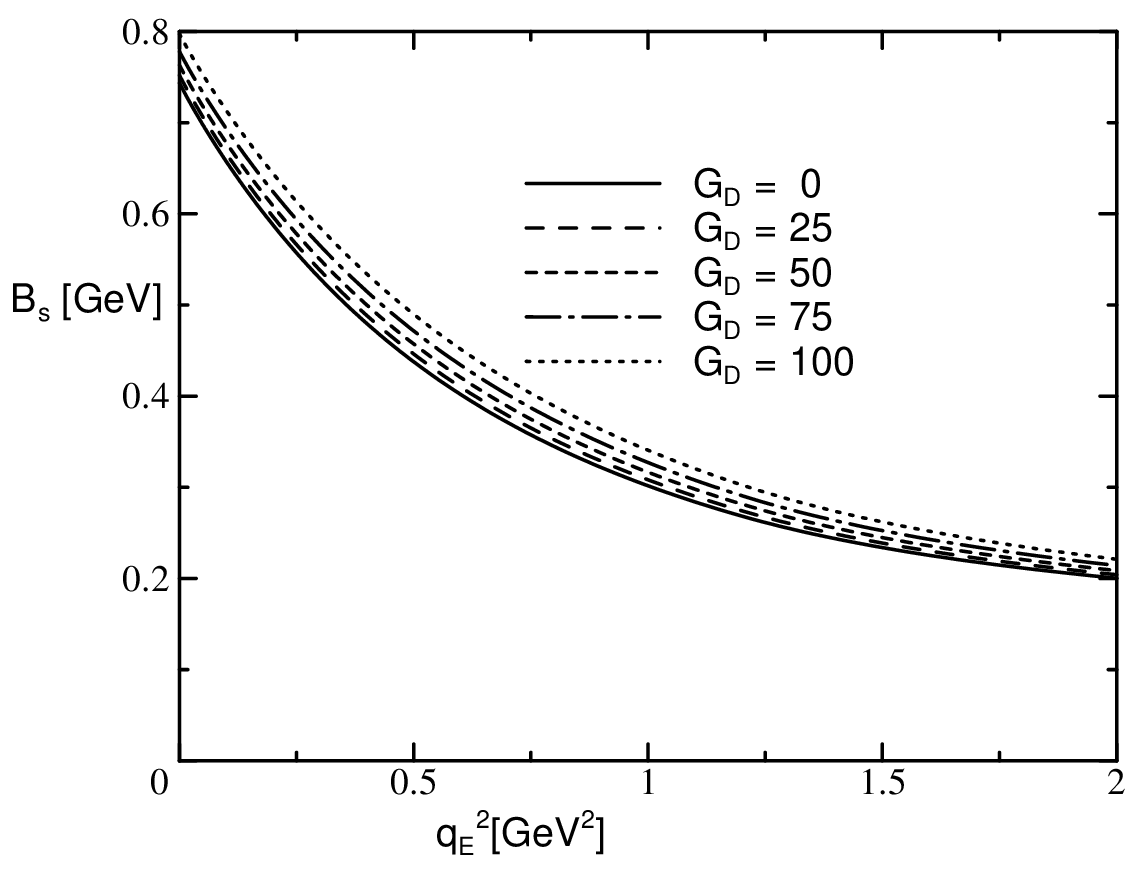} }
  \caption{$q_E^2$ dependences of the mass function of the q quark $B_s$ with 
           $G_D=0, 25, 50, 75,$ $100 [\mbox{GeV}^{-5}]$}
  \label{FIG:Bs}
\end{figure}

Let us now discuss the solutions of the SD equation. 
Our numerical results of the running quark masses are 
shown in Fig.{$\ref{FIG:Bq}$} and {$\ref{FIG:Bs}$}. 
The values of the quark condensates 
and mass function at $P_E^2 = 0$ are 
given  in the Table $\ref{TBL:sdResult}$. 

\begin{table}[tbp]
\begin{center}
\begin{tabular}{|c|cccc|}  \hline
$G_D [\mbox{GeV}^5]$&$B_q(0) [\GeV] $&$B_s(0)[\GeV]$
&$-\langle\bar{q}q\rangle_R^{1/3}[\mbox{GeV}]$
&$-\langle\bar{s}s\rangle_R^{1/3}[\mbox{GeV}]$\\\hline
0  &0.524  &0.744  &$ 0.225$ &$ 0.052$ \\
25 &0.551  &0.752  &$ 0.236$ &$ 0.088$ \\
50 &0.582  &0.763  &$ 0.248$ &$ 0.122$ \\
75 &0.616  &0.778  &$ 0.259$ &$ 0.146$ \\
100 &0.655  &0.797  &$ 0.270$ &$ 0.165$ \\
\hline
\end{tabular} 
\end{center}  
 \caption{Dependences of the mass function at $q_E^2=0$ and the quark
          condensates on the strength of the KMT interaction.}
 \label{TBL:sdResult}
\end{table}    

\begin{figure}[tbh]
  \centerline{ \epsfysize= 6cm \epsfbox {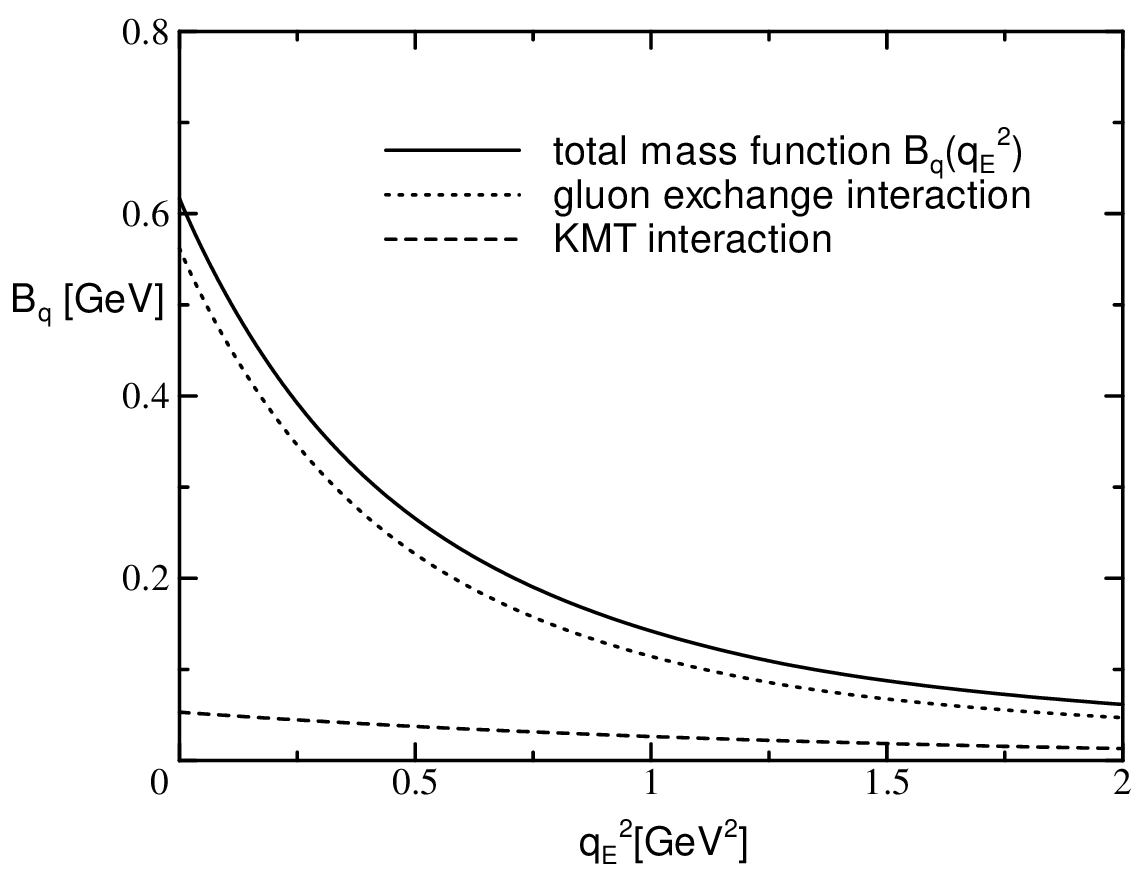} }
  \caption{Effects of the gluon exchange interaction and the KMT interaction
   to the q quark mass function at $G_D = 75 [\GeV^{-5}]$}
  \label{FIG:sdRatio_q}
\end{figure}

\begin{figure}[tbh]
  \centerline{ \epsfysize= 6cm \epsfbox {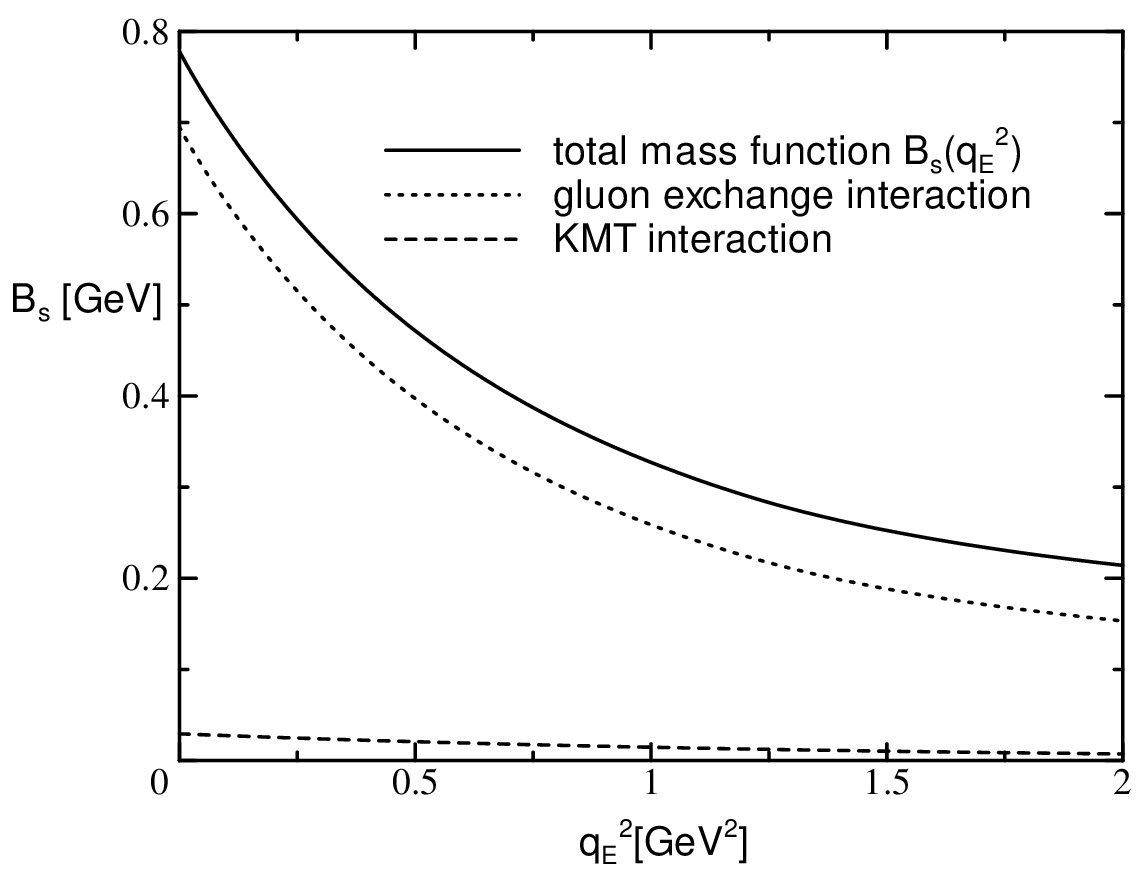} }
  \caption{Effects of the gluon exchange interaction and the KMT interaction to the s quark mass function at $G_D = 75 [\GeV^{-5}]$}
  \label{FIG:sdRatio_s}
\end{figure}

One sees that the mass function  depends  weakly 
on $G_D$. When $G_D$ increases from 0 to 75 $[\GeV^{-5}]$, $B_q(0)$ 
increases only by 15\%. 
It becomes clearer  by comparing the 
effects in the SD equation of the one gluon exchange term 
with that of the KMT term. 
It is shown in Figs. $\ref{FIG:sdRatio_q}$ 
and $\ref{FIG:sdRatio_s}$.


\begin{figure}[t]
  \centerline{ \epsfysize= 6cm \epsfbox {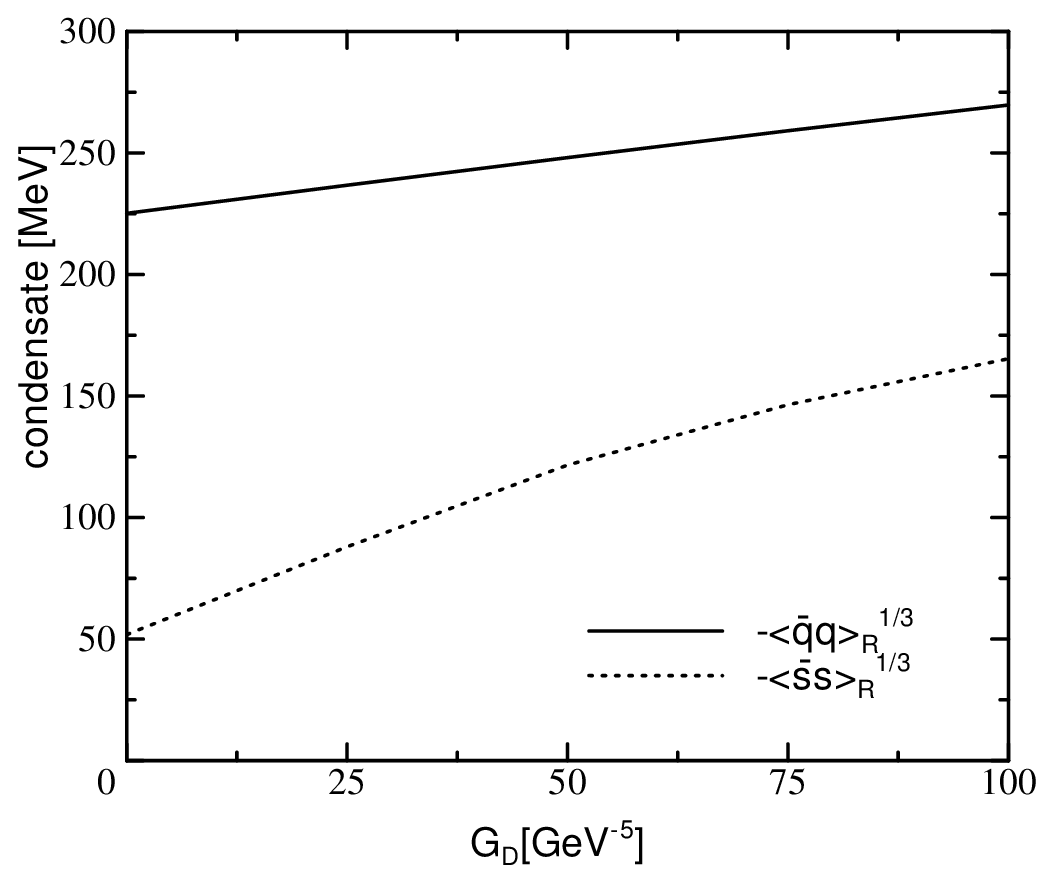} }
  \caption{Dependence of the quark condensates on the strength 
           of the KMT interaction.}
  \label{FIG:conden}
\end{figure}

The dependence of the condensates of $u,d$ quark on $G_D$ is also small as 
shown in Fig. $\ref{FIG:conden}$. 
When $G_D$ increases from 0 to 75 $[\GeV^{-5}]$, 
$- \la \bar{q}q \ra_R^{1/3}$ changes by 13 \%. 
But at the same time $\la \bar{q}q \ra_R$ 
changes by 24\%, 
and indicate that 
the effects of the KMT interaction is not so small. 
The dependence of the condensates of $s$ quark on $G_D$ is rather large. 
When $G_D$ increases from 0 to 75 $[\GeV^{-5}]$, 
$- \la \bar{s}s \ra_R^{1/3}$ changes by 64 \%. 
%


%

These results clearly differ from the result of NJL model. 
In the NJL model, the effect of the KMT interaction 
in the dynamical quark mass 
are very large,  $G_D\la \bar{s}s \ra/G_S = 0.44$ suggests 
that the contribution from the KMT interaction  
 to the dynamical quark mass is 
 44\% of that of the four quark interaction. 
This difference will be examined later.

\section{Solution of the BS equation of  the Pseudo Scalar mesons}

Let us now turn to the discussion of the solutions of the BS equation. 
Our numerical results for the pion are summarized in Table \ref{TBL:result_pi}.
From these results, it is shown that the pion mass
 is not so sensitive to $G_D$, which suggest 
that 
the increase of dynamical quark mass and 
of the attractive KMT interaction tend to cancel with each other. 
The change of decay constant is also small.  It is 7\% and this 
result is consistent with the change of the condensate. 
The R value defined in Eq.($\ref{Eq:RValue}$) is an indicator which shows whether the Ward identity 
of the axial vector current is satisfied  on the mass shell. 
We consider that the deviation  of R from the unity 
indicates  error in  the extrapolation from Euclid $(p_E^2 > 0)$ 
to Minkowski $(p_E^2 < 0)$ region. 
From the Table $\ref{TBL:result_pi}$, one 
sees that the deviation is about $\sim 6\%$ 
and thus extrapolation error for $M_{\pi}^2$ and $f_{\pi}$ 
is of this order.

\begin{table}[tbh]
\begin{center}
\begin{tabular}{|c|ccc|}  \hline
$G_D[\GeV^{-5}]$& $M_{\pi}[\MeV]$ & $f_{\pi}[\MeV]$ &R \\
\hline
0  &126 &84 &1.00 \\
25 &130 &88 &1.02 \\
50 &133 &92 &1.03 \\
75 &136 &95 &1.05 \\
100&139 &99 &1.06 \\
\hline
\end{tabular} 
\end{center}  
 \caption{Dependences of the solution of the pion BS equation 
          on the strength of the KMT interaction.}
 \label{TBL:result_pi}
\end{table}    

\begin{figure}[tb]
  \centerline{ \epsfysize= 12cm \epsfbox{ 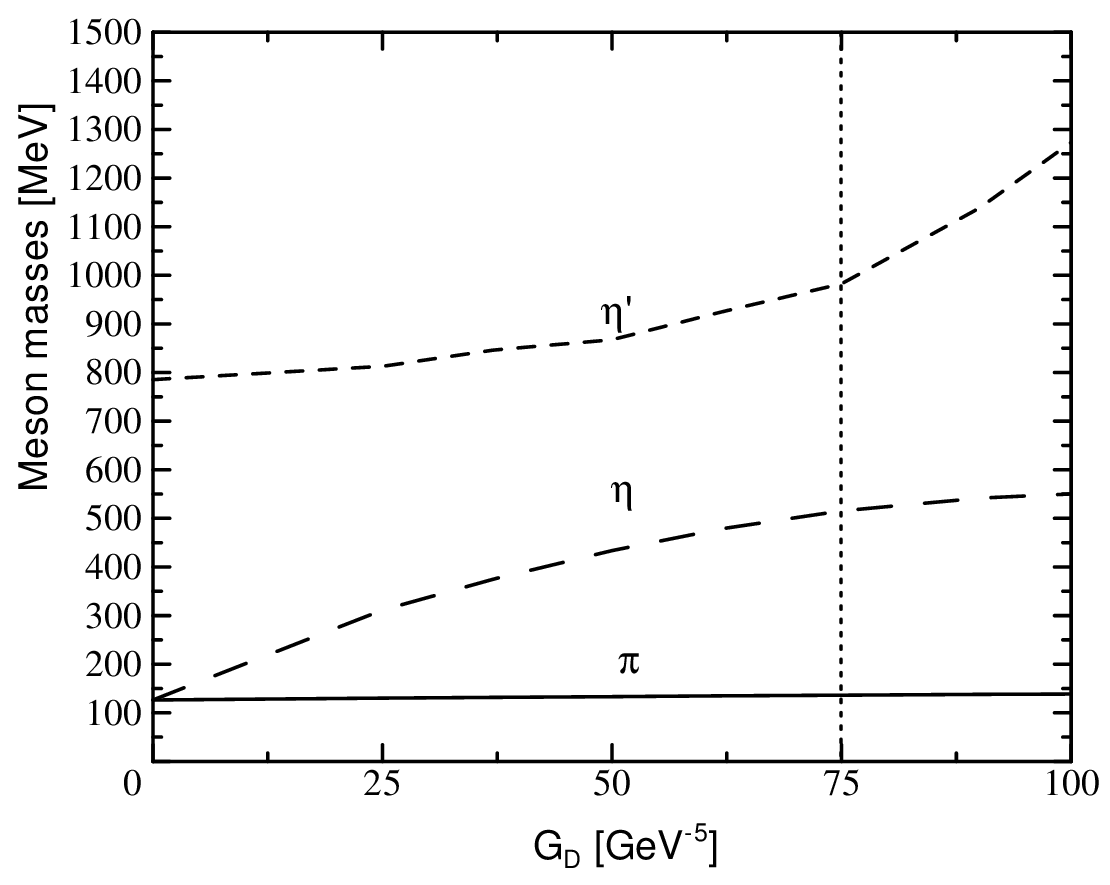} }
  \caption{Dependence of the mass spectrum of the 
           $\pi$, $\eta$ and $\eta'$ mesons 
           on the strength of the KMT interaction.}
  \label{FIG:psMeson}
\end{figure}

The masses and decay constants  of the $\eta$ and $\eta'$ are 
shown in Tables \ref{TBL:result_eta}, \ref{TBL:result_etp}
and Fig. ($\ref{FIG:psMeson}$). 
Since the BS equation is homogeneous, the sign of the 
BS amplitudes, and therefore the decay constants, 
is undetermined. We choose the sign so that $f_8$ is positive for $\eta$, 
and $f_0$ is positive for $\eta'$. 

The masses of $\eta$ and $\eta'$ mesons and their decay constants 
 depend strongly on the $G_D$. 
Especially, the $\eta'$ meson mass  seems to be sensitive as 
is expected. 
This is in contrast to the result for the pion. 
The $U_A(1)$ breaking effects are large in the 
$\eta$ and $\eta'$ sector. 
The overall behavior of the mass spectrum 
agrees well with the NJL result. \cite{HK1994,TNO1997}

\begin{table}[tbp]
\begin{center}
\begin{tabular}{|c|ccccc|}  \hline
$G_D[\GeV^{-5}]$& $M_{\eta}[\MeV]$ & $f_{\eta}^8 [\MeV]$ &$f_{\eta}^0 [\MeV]$
 &$R^8$ &mixing angle[deg] \\
\hline
0  &134 &49   &69   &1.00  &-54.7 \\
25 &311 &61   &65   &1.06  &-46.8 \\
50 &409 &78   &53   &1.08  &-34.2 \\
75 &447 &95   &35   &1.08  &-20.0 \\
100&448 &106  &16   &1.03  &-8.6  \\
\hline
\end{tabular} 
\end{center}  
 \caption{Dependences of the solution of the $\eta$ meson solution 
          of the coupled channel BS equation 
          on the strength of the KMT interaction.}
 \label{TBL:result_eta}
\end{table}    
\begin{table}[tbp]
\begin{center}
\begin{tabular}{|c|ccccc|}  \hline
$G_D[\GeV^{-5}]$& $M_{\eta'}[\MeV]$ & $f_{\eta'}^8 [\MeV]$ 
 &$f_{\eta'}^0 [\MeV]$
 &$R^8$ &mixing angle[deg] \\
\hline
0   & 785  & -91 & 74  & 1.55  & -54.8 \\
25  & 846  & -71 & 91  & 1.69  & -35.9 \\
50  & 867  & -54 & 103 & 1.71  & -34.2 \\
75  & 982  & -28 & 109 & 1.73  & -26.2 \\
100 & 1273 & -11 & 82  & 2.05  & -6.6  \\
\hline
\end{tabular} 
\end{center}  
 \caption{Dependences of the solution of the $\eta'$ meson solution 
          of the coupled channel BS equation 
          on the strength of the KMT interaction.}
 \label{TBL:result_etp}
\end{table}    

\begin{figure}[h]
  \centerline{ \epsfysize= 10cm \epsfbox{ 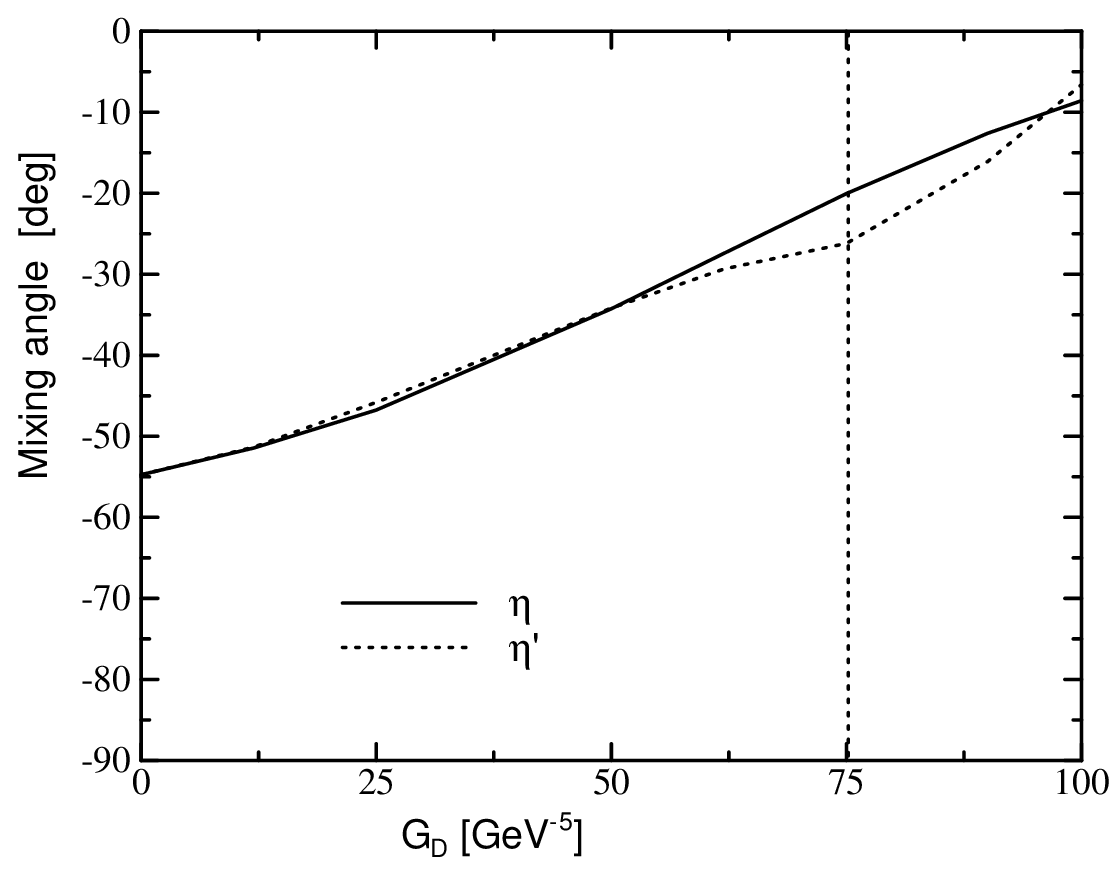} }
  \caption{Dependence of the mixing angle of the $\eta$ and $\eta'$ meson on the strength of the KMT interaction}
  \label{FIG:mixingEta}
\end{figure}

The $\eta - \eta'$ mixing angles are defined 
in terms of  the decay constants as 
\begin{eqnarray}
\frac{- f^{\eta}_0}{f^{\eta}_8}  = \tan \theta_{\eta} \\
\frac{f^{\eta'}_8}{f^{\eta'}_0}  = \tan \theta_{\eta'}.
\end{eqnarray}
The results are presented in Tables \ref{TBL:result_eta}, \ref{TBL:result_etp} 
and Fig. ($\ref{FIG:mixingEta}$). 
In the limit of $G_D = 0$, since 
there is no flavor mixing, $\eta$  is pure 
$\frac{1}{\sqrt{2}}(u\bar{u}+d\bar{d})$ state 
and 
$\eta'$ is $s\bar{s}$ state.
They are the ideally 
mixed states, i.e., 
$\theta = \arctan(-\sqrt{2}) = -54.7\,^{\circ}$. 
In this limit,
as $G_D$ becomes larger, the mixing angle approaches to $0^{\circ}$ 
i.e., the $SU(3)$ limit. 
At that time, 
$\eta$ approaches to the pure octet 
$\frac{1}{\sqrt{6}}(u\bar{u} +d\bar{d} -2s\bar{s})$ state 
and 
$\eta'$ approach to the pure singlet 
$\frac{1}{\sqrt{3}}(u\bar{u} +d\bar{d}+s\bar{s})$
 state. 
At $G_D = 75 [\GeV^{-5}]$, we obtain $\theta_{\eta} = -20.0\,^{\circ}$ and 
$\theta_{\eta'} = -26.2\,^{\circ}$. 
These large changes of the mixing angle and the mass spectrum at 
$G_D=75 [\GeV^{-5}]$ indicate strong $U_A(1)$ symmetry breaking 
effect. 

Finally we check the extrapolation ambiguities in 
the $\eta$ and $\eta'$ meson masses. 
The R values for the $\eta$ meson are as good as the pion case. 
On the other hand, the deviation  for the  $\eta'$ meson is as much as  73\%.
%
As the $\eta'$ mass is much larger than the $\pi$ and $\eta$ 
masses, 
the extrapolation from the Euclid region 
to the on shell momentum brings large uncertainty. 
It is shown in Fig. $\ref{FIG:R_etp_090}$. 
\begin{figure}[tb]
  \centerline{ \epsfysize= 8cm \epsfbox{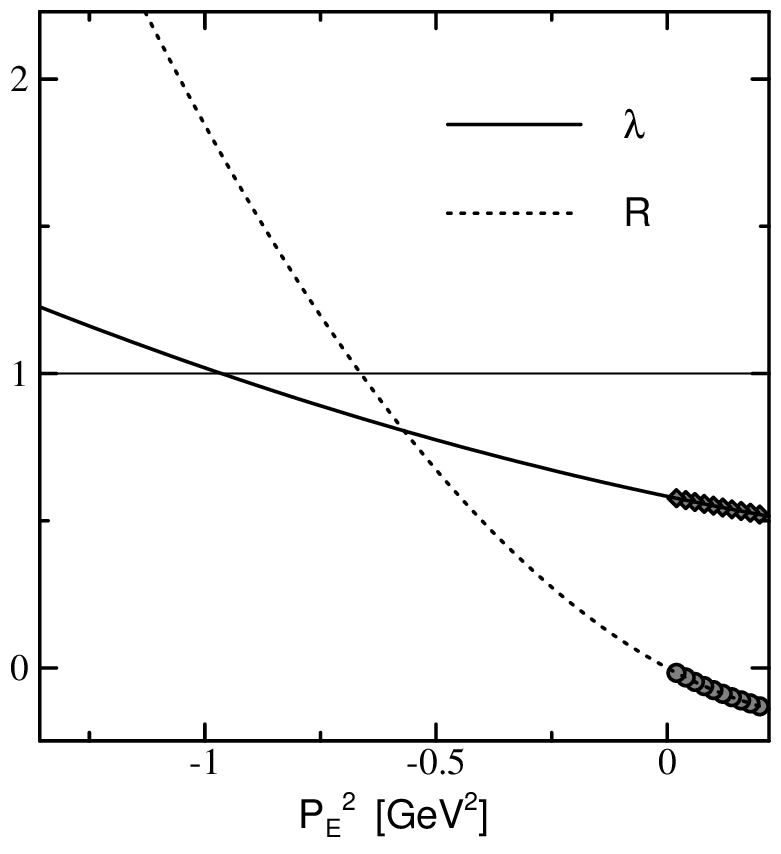} }
  \caption{The extraporation of the eigenvalue $\lambda$ of the 
  BS equation and R value. 
  The solid line shows the extraporation of the $\lambda$ 
  and the doted line shows the extraporation of the $R$. 
   }
  \caption{Extrapolation of the eigenvalue and the $R$ for $\eta'$ meson 
  with $G_D = 75 [\GeV^{-5}]$.  }
  \label{FIG:R_etp_090}
\end{figure}
The obtained mass from the condition $R = 1$ is 
$814 [\MeV]$ and the  mass obtained from 
the condition $\lambda = 1$, where $\lambda$ is the eigenvalue 
of the BS equation,  is $982 [\MeV]$. 
This difference  of the masses may show the extrapolation uncertainty. 
Since the dependence of the decay constants $f^{0}_{\eta'}$ and $f^{8}_{\eta'}$
on the momentum $P_E^2$  is small, the uncertainty  of the decay constants 
is smaller than that of the masses. 
Those are shown in Fig. $\ref{FIG:R_etp_090}$. 
\begin{figure}[tb]
  \centerline{ \epsfysize= 8cm \epsfbox{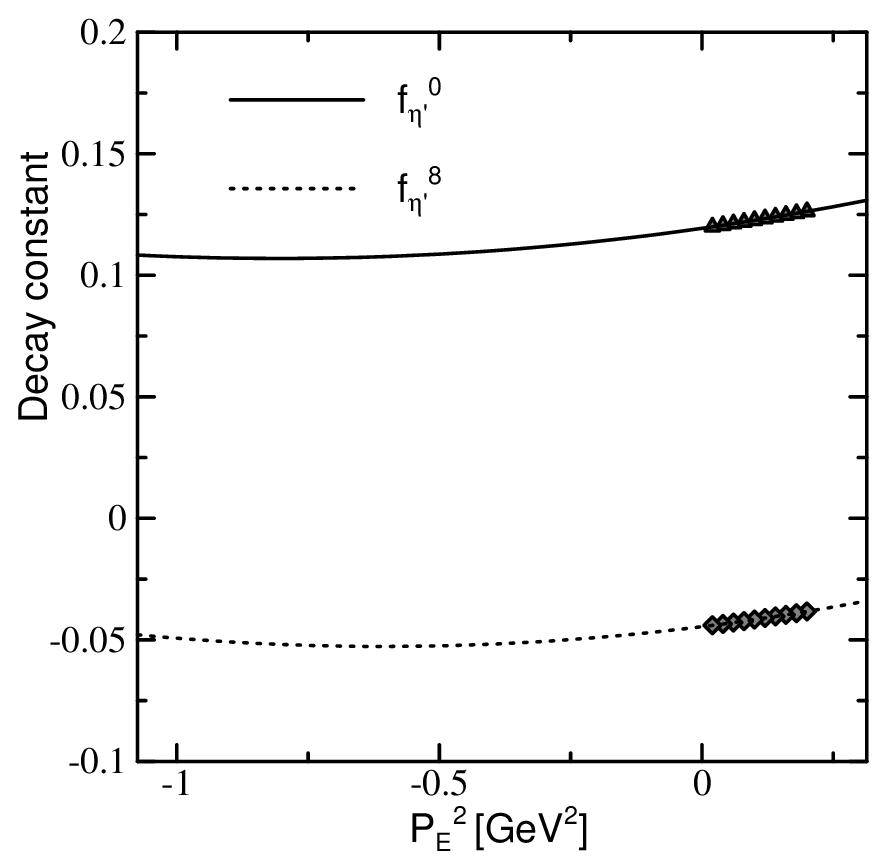} }
  \caption{Extraporation of the $f^{0}_{\eta'}$ and $f^{8}_{\eta'}$ 
   with $G_D = 75 [\GeV^{-5}]$.}
  \label{FIG:R_etp_090}
\end{figure}
We think that this ambiguity from the 
extrapolation is unavoidable in the 
current approach.


\section{Solution of the BS equation of the Scalar Mesons}

\begin{figure}[tb]
  \centerline{ \epsfysize= 12cm \epsfbox{ 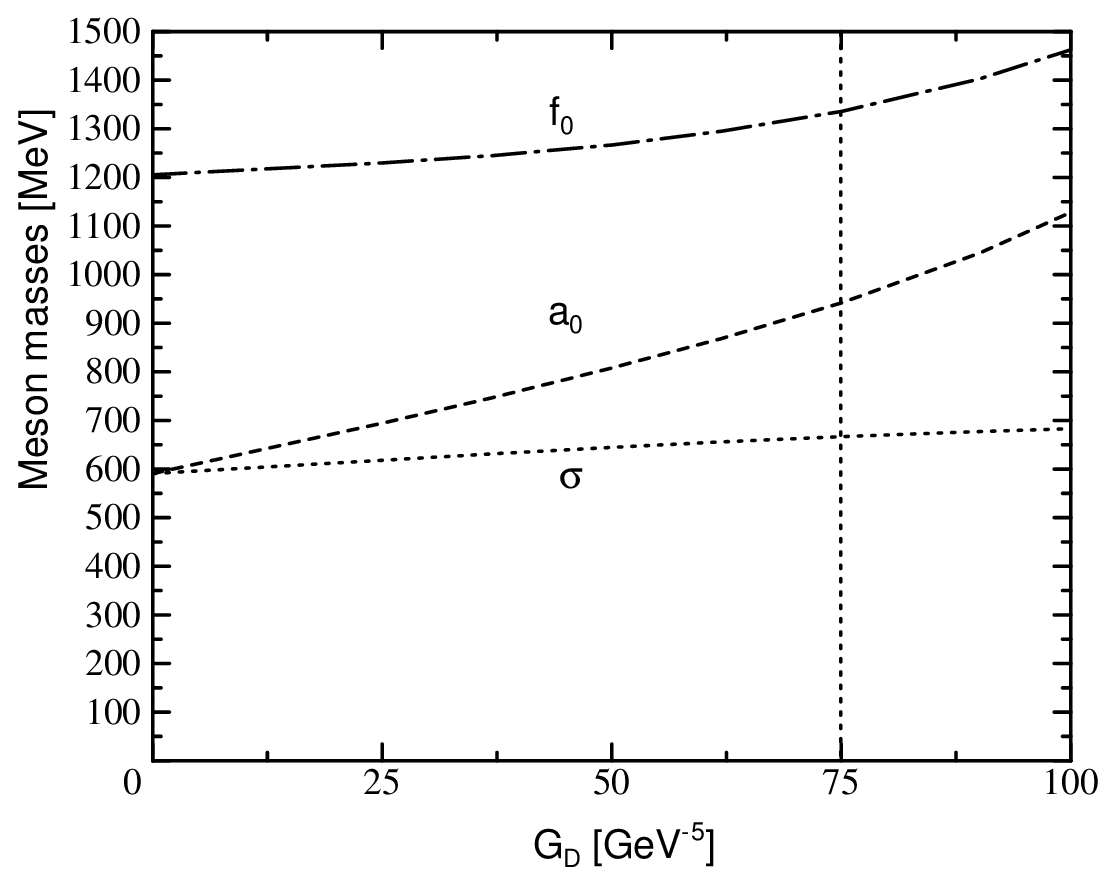} }
  \caption{Dependence of the mass spectrum of the 
           $a_0$, $\sigma$ and $f_0$ mesons 
           on the strength of the KMT interaction.}
  \label{FIG:sMeson}
\end{figure}

In this section, we present the results of 
the ILA calculation for the scalar mesons. 
Note that the parameters of the present approach are completely 
fixed in the study of  the pseudoscalar mesons. 
The results for the  $a_0$, $\sigma$, $f_0$ mesons are 
summarized in Tables $\ref{TBL:result_a0}$, $\ref{TBL:result_sig}$ and 
$\ref{TBL:result_f0}$. The dependence of the mass spectrum 
on the strength, $G_D$ of the KMT interaction 
is shown in Fig. $\ref{FIG:sMeson}$. 

First, the dependencies  of the masses of $\sigma$ and $a_0$ 
on $G_D$ 
look qualitatively  same as the NJL results shown in Fig.($\ref{fig:mass}$). 
In the NJL calculation, the parameters are chosen so as to 
reproduce the $M_{\pi}$ and $f_{\pi}$ at each $G_D$. 
In contrast, in the ILA approach we change   $G_D$ independently. 
However, since $M_{\pi}$ and $f_{\pi}$ depend weakly on 
$G_D$, the results of ILA 
show similar behavior as those of NJL. 
We note that the mass of $a_0$ grows as $G_D$ increases, while 
the $\sigma$ mass is fairly constant. 
The behavior of the $f_0$ mass differs from the that of NJL model. 
As noted in Chapter 3, the mass of the $f_0$ solution in the NJL model is 
much higher than the threshold, $2M_q$ and the momentum cutoff,$\lambda$. 
The $f_0$ mass in the NJL result is not so reliable.

The $\sigma$ meson mass is predicted as about $670 [\MeV]$. 
This rather small $\sigma$ meson mass is interesting. 
In the case of the NJL model, 
the $\sigma$ meson mass is determined to be close to 
twice of the 
dynamical quark mass. 
On the other hand, in the ILA approach 
the value of the mass function at $P_E^2$, $B_q(p_E^2=0)$ 
is about $616 [\MeV]$, 
which is comparable to the $\sigma$ meson mass. 
Although there are such differences, 
the properties of 
the physical observables agree in both 
the calculations. 
This feature is very interesting.

The dependence of the $\sigma$ meson mass on the $G_D$ is 
small. This means that the effects of the $s\bar{s}$ mixing 
and the KMT interaction are balanced. 
On the other hand, in the case of $a_0$, 
the flavor content does not change and 
only the 
repulsion of the KMT is effective. 
Thus the $a_0$ mass monotonically grows as $G_D$ increases. 
We obtain $M_{a_0} = 942 [\MeV]$. This result is comparable 
to the experimental value $984.8 \pm 1.4 [\MeV]$. 
We obtain a significant mass splitting between the $\sigma$ and $a_0$,
 about $272[\MeV]$. 
We conclude that the observed $\sigma-a_0$ mass splitting 
 can be explained as the 
$U_A(1)$ symmetry breaking effects.

The obtained mass of $f_0$ is larger than the 
mass of $a_0$ by about $394\MeV$ and they 
do not become degenerate even if we change the strength of the 
KMT interaction in our parameter range. 
Therefore $f_0(980)$  may not be identified with the $q\bar{q}$ state.
To reproduce $f_0(980)$ and the higher $I=0$ scalar meson states, 
 it may be needed to treat mixings of multi-quark states. 

\begin{table}[tbp]
\begin{center}
\begin{tabular}{|c|c|}  \hline
$G_D[\GeV^{-5}]$& $M_{a_0}[\MeV]$  \\
\hline
0  & 591 \\
25 & 694 \\
50 & 808 \\
75 & 942 \\
100& 1128\\
\hline
\end{tabular}
\end{center} 
 \caption{Dependences of the solution of the $a_0$ BS equation 
          on the strength of the KMT interaction.}
 \label{TBL:result_a0}
\end{table}
\begin{table}[tbp]
\begin{center}
\begin{tabular}{|c|cccc|}  \hline
$G_D[\GeV^{-5}]$& $M_{\sigma}[\MeV]$
 &$S^0$ &$S^8$  &mixing angle[deg] \\
\hline
0  & 591  & 12.9 & 9.09 & -54.7 \\
25 & 618  & 13.5 & 8.40 & -58.1  \\ 
50 & 645  & 14.4 & 7.51 & -62.5 \\ 
75 & 667  & 16.0 & 6.47 & -68.0 \\ 
100& 683  & 18.1 & 5.25 & -73.8  \\ 
\hline
\end{tabular} 
\end{center}  
 \caption{Dependences of the solution of the $\sigma$meson solution 
          of the coupled channel BS equation 
          on the strength of the KMT interaction.}
 \label{TBL:result_sig}
\end{table}    
\begin{table}[tbp]
\begin{center}
\begin{tabular}{|c|cccc|}  \hline
$G_D[\GeV^{-5}]$& $M_{f_0}[\MeV]$
 &$S^0$ &$S^8$  &mixing angle[deg] \\
\hline
0  & 1205 &-21.1   & 14.9  & -54.7  \\
25 & 1230 &-20.8   & 7.4  &  -70.5 \\ 
50 & 1267 &-28.4   & 3.1  &  -83.7 \\ 
75 & 1336 &-35.5   & 3.8  &  -83.9 \\ 
100& 1462 &-45.9   & 10.3  & -77.3  \\ 
\hline
\end{tabular} 
\end{center}  
 \caption{Dependences of the solution of the $f_0$ meson solution 
          of the coupled channel BS equation 
          on the strength of the KMT interaction.}
 \label{TBL:result_f0}
\end{table}    

\begin{figure}[tb]
  \centerline{ \epsfysize= 8cm \epsfbox{ 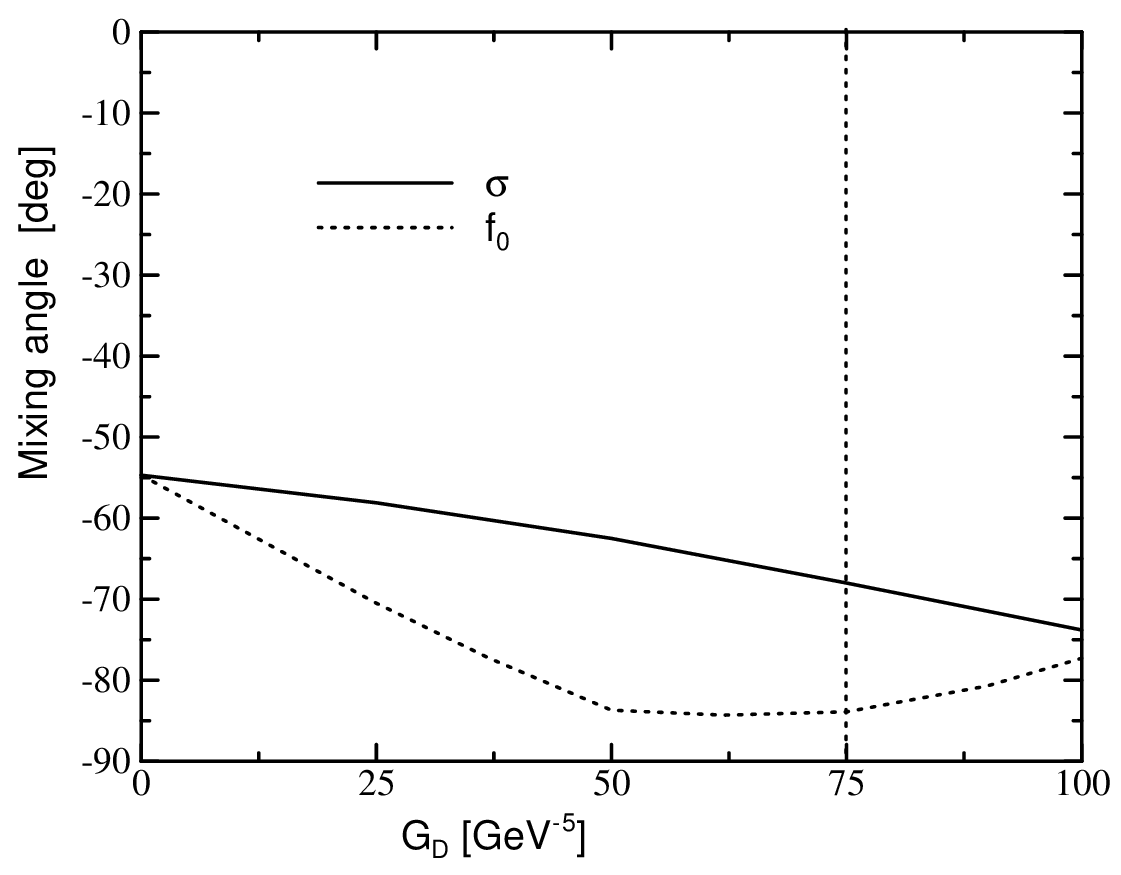} }
  \caption{Dependence of the mixing angle of the $\eta$ and $\eta'$ meson on the strength of the KMT interaction.}
  \label{FIG:mixing_sig}
\end{figure}

Next, we consider the mixing angles. 
We introduce the matrix elements $S^8$ and $S^0$ which are defined by 
\begin{eqnarray}
&&S^a =\int d^4x \langle 0 | \bar{q} \frac{\lambda^8}{2} q (x)
| \mbox{scalar meson state} \rangle \\
&&= \tr \left[\bar{\chi}^R(0;P)\frac{\lambda^{a}}{2} \right]
\end{eqnarray}
Since these S values extract the particular flavor 
component of $\phi_S$ which is 
the main component of the BS amplitude at the origin, 
we employ $S^8$ and $S^0$ to determine the 
ratio of the octet and 
the singlet components. 
Accordingly we define the mixing angles 
of the scalar mesons as 
\begin{eqnarray}
&&\tan \theta_{\sigma} = - \frac{S_0^{\sigma}}{S_8^{\sigma}} \\
&&\tan \theta_{f_0} = \frac{S_8^{f_0}}{S_0^{f_0}}. 
\end{eqnarray}
The results are summarized 
in Table $\ref{TBL:result_a0}$, $\ref{TBL:result_sig}$ and 
Fig $\ref{FIG:mixing_sig}$. 

The mixing angle of $\sigma$ approaches $-90^\circ$, where 
$\sigma$ becomes the purely flavor singlet state: 
$\frac{1}{\sqrt{3}}(u\bar{u} + d\bar{d} + s\bar{s})$.
The obtained angle at $G_D=75[\GeV^{-5}]$ is $-68.0^\circ$ and is 
slightly smaller than the result of the 
NJL model, $-77.3^\circ$. This angle corresponds to 
about $5\%$ mixing of the strangeness  component 
in $\sigma$.

Finally ,
the comparison of the scalar and pseudoscalar meson spectra
is shown in Fig. $(\ref{FIG:spectrum})$
\begin{figure}[tb]
  \centerline{ \epsfysize= 12cm \epsfbox{ 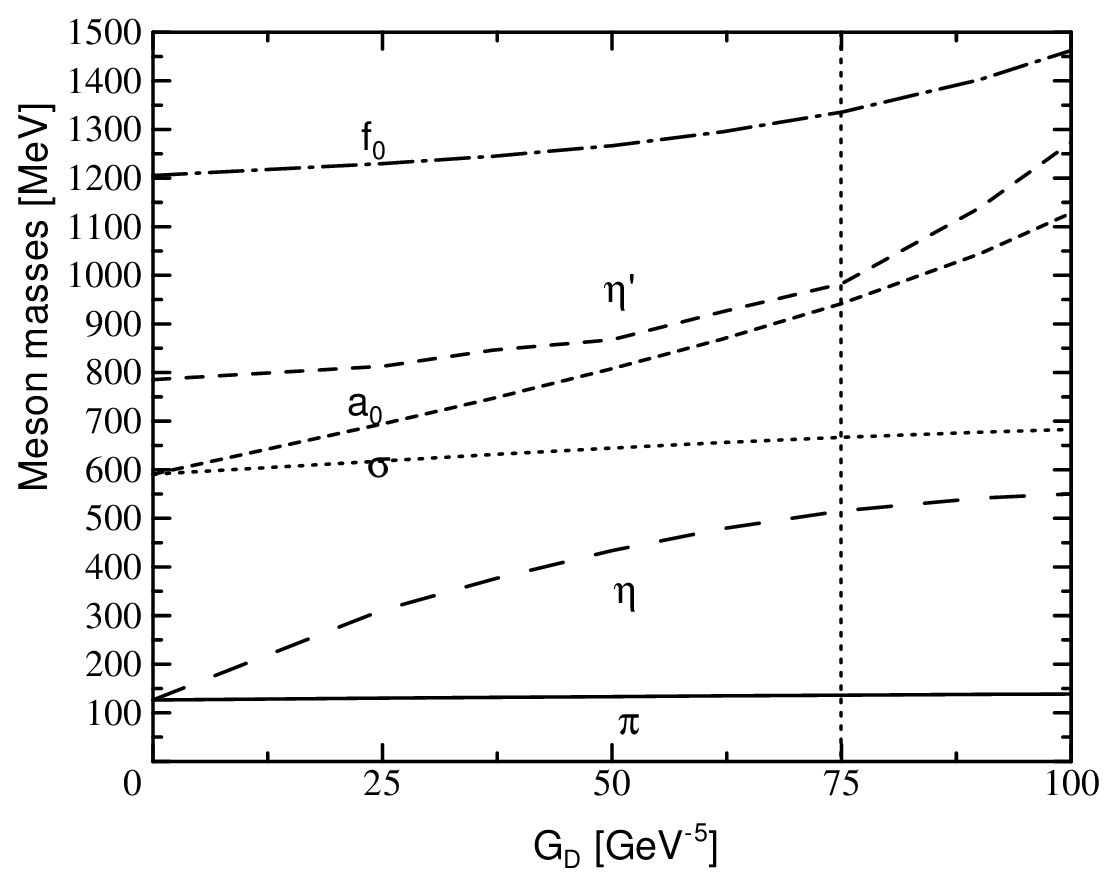} }
  \caption{Dependence of the mass spectrum of the 
           $\pi$, $\eta$, $\eta'$, $a_0$, $\sigma$ and $f_0$ mesons 
           on the strength of the KMT interaction.}
  \label{FIG:spectrum}
\end{figure}
%

\section{Solution of the BS equation of the Strange Mesons}

In this section, we 
calculate the masses of 
the strange pseudoscalar  meson $K$ 
and the strange scalar meson $K_0^*$. 
Here we employ a crude approximation in order to 
avoid the technical difficulty. 
As these mesons consists of the a strange quark and a  
nonstrange quark. 
Then 
the BS equation of the ILA approach becomes extremely complicated. 
Instead of treating the asymmetric  BS equation, 
we solve the symmetric BS equation for the 
quarks whose mass is $77.25 [\MeV]$.  This mass is the average of 
$m_q$ and $m_s$. 
The results are summarized in Tables $\ref{TBL:result_Kon}$,
$\ref{TBL:result_kappa}$ and Fig.($\ref{FIG:spectrum}$). 

\begin{table}[btp]
\begin{center}
\begin{tabular}{|c|c|}  \hline
$G_D[\GeV^{-5}]$& $M_{K}[\MeV]$ \\
\hline
0 & 547 \\
25 & 546 \\
50 & 544 \\
75 & 538 \\
100 & 530 \\
\hline
\end{tabular} 
\end{center}  
 \caption{Dependences of the solution of the $K$ meson BS equation 
          on the strength of the KMT interaction.}
 \label{TBL:result_Kon}
\end{table}
\begin{table}[btp]
\begin{center}
\begin{tabular}{|c|c|}  \hline
$G_D[\GeV^{-5}]$& $M_{K_0^*}[\MeV]$ \\
\hline
0 &  946 \\
25 & 1040  \\
50 & 1158  \\
75 & 1321  \\
100 & 1542  \\
\hline
\end{tabular} 
\end{center}  
 \caption{Dependences of the solution of the $K_0^*$ meson BS equation 
          on the strength of the KMT interaction.}
 \label{TBL:result_kappa}
\end{table}
%


Again the $K$ mass depends little 
on $G_D$ as in the case of $\pi$
. 
On the other hand, the $K_0^*$ meson mass 
increases due to the repulsion of the KMT interaction.

Although these results are based on the symmetric BS equations, 
the obtained $K$ mass, $494 \sim 498 [\MeV]$, 
reproduces the observed value and 
$G_D$ dependence agrees with that of the NJL.

%
\begin{figure}[tb]
  \centerline{ \epsfysize= 12cm \epsfbox{ 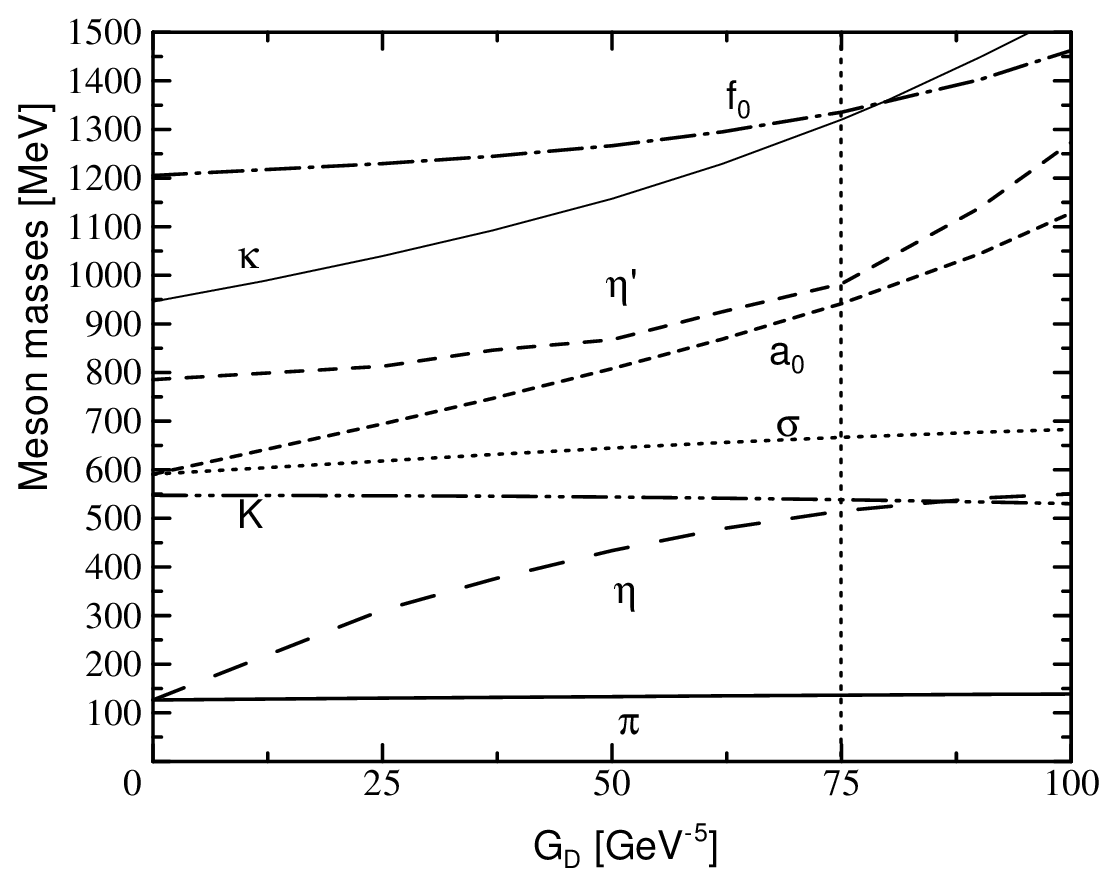} }
  \caption{Dependence of the mass spectrum 
           of the scalar and pseudoscalar meson nonet
           on the strength of the KMT interaction.}
  \label{FIG:spectrum}
\end{figure}
One sees that the slopes of the $a_0$ mass and $K_0^*$ mass 
are nearly equal and therefore  it is clear that 
the $K_0^*$ mass does not become smaller than 
the $a_0$ mass in this approach. 
Since the NJL also gives the similar  result,
we conclude that 
 the reversal of 
the $\kappa(700-900)$ mass and the $a_0(980)$ mass can not be 
explained by the $U_A(1)$ symmetry breaking. 
Furthermore, our $K_0^*$ mass is smaller than the next excited state 
$K_0^{*}(1430)$ by about $110 \MeV$. 
Consequently,  the light $I = 1/2$ scalar mesons 
may not be explained without considering mixings 
of the multi-quark states.

\chapter{Conclusion}
We have studied the roles of the $U_A(1)$ breaking 
interaction of QCD in the spectrum of the
 light scalar nonet mesons using 
the extended NJL model and 
the ILA of QCD. 
We have first analyzed qualitative properties of 
the $U_A(1)$ breaking interaction 
in the extended NJL model, and next 
using the ILA approach we have carried out 
quantitative study. 
Since these approaches are consistent with 
chiral symmetry, 
the scalar and pseudoscalar mesons can be 
regarded  as chiral partners. 
We choose parameters to
 reproduce the masses and decay constants of the 
pseudoscalar nonet mesons 
and apply those to the scalar nonet mesons.

The extended NJL model is the SU(3) NJL model 
supplemented with the Kobayashi-Maskawa-'t Hooft (KMT) 
interaction, which causes the $U_A(1)$ breaking. 
The strength of the KMT interaction is chosen  so as to 
explain the electromagnetic decays of the $\eta$ meson. 
To study  the role of the $U_A(1)$ breaking, 
we study the mass spectrum and the 
mixing angle of the scalar mesons as functions of  
the strength of the KMT interaction. 
As a result,  we observe  that the  $\sigma$ meson mass 
changes little, which suggests that 
the increase of the dynamical quark mass and 
the attractive KMT interaction cancel with each other. 
On the other hand, 
the mass of $a_0$ increases monotonically due to the 
repulsive KMT interaction. 
This $\sigma-a_0$ mass splitting amounts to a 
 $150$ MeV and is slightly smaller than the 
current experimental data. 
A possible reason that the $a_0$ mass is not large enough is the large
unphysical $q\qbar$ decay width 
which is an artifact of this model.
We have also found that the strangeness content in 
the $\sigma$ meson is about $15\%$.

The ILA of QCD is 
 an approximation that is consistent with chiral 
symmetry and consists of the rainbow approximation of the 
Schwinger-Dyson equation and the ladder approximation 
of the Bethe-Salpeter equation. 
Using this approach, we analyze 
the scalar meson spectrum 
quantitatively. 
As we do in  the extended NJL model,  
we choose the parameters to 
reproduce physical observables  of the 
pseudoscalar mesons. 

We have obtained that the $\sigma$ meson 
mass comes around $600 \sim 700[\MeV]$, 
which is much lower than the value expected 
from the effective quark mass. 
This suggests that the $\sigma$ meson 
is special as the  chiral partner of the pion. 
We obtain the $\sigma-a_0$ mass splitting of about 
$272[\MeV]$ and the strangeness 
 content in 
the $\sigma$ meson is about $5\%$. 
It will be interesting  to 
check this $s\bar{s}$ mixing experimentally 
by using, for instance, the 
radiative  $J/\psi$ decays in near future. 

The obtained mass of $f_0$ is larger than the 
mass of $a_0$ by about $394\MeV$ and they 
do not become degenerate even if we change the strength of the 
KMT interaction in our parameter range. 
Therefore $f_0(980)$  may not be identified with the $q\bar{q}$ state.
To reproduce $f_0(980)$ and the higher $I=0$ scalar meson states, 
 it may be needed to treat mixings of multi-quark states. 

The masses of  the strange mesons have been calculated 
in an approximation. We have found that the $K_0^*$ mass is 
larger than the $a_0$ mass by about $230 [\MeV]$. 
The masses of $a_0$ and $K_0^*$ mesons 
increase by almost the same rate as $G_D$ is increased. 
Therefore it seems that the reversal of 
the $\kappa(700-900)$ mass and the $a_0(980)$ mass can not be 
explained by the $U_A(1)$ symmetry breaking. 

Furthermore, our $K_0^*$ mass is smaller than the next excited state 
$K_0^{*}(1430)$ by about $379 \MeV$. 
Consequently,  the light $I = 1/2$ scalar mesons 
may not be explained without considering mixings 
of multi-quark states.

In conclusion, 
we have shown that 
the spectra of the 
pseudoscalar meson nonet and 
the $\sigma$ and $a_0$ masses 
can be reproduced by the extended 
Nambu-Jona-Lasinio 
model 
and improved ladder approximation of QCD 
with $U_A(1)$ breaking interaction. 
We have found that the 
observed spectrum is consistent with 
the strong $U_A(1)$ breaking interaction. 
Furthermore we have shown that $f_0(980)$ and 
$\kappa(700-900)$ can not be reproduced in 
our picture.  
It may be needed to treat mixings of multi-quark states.

For the more detailed  analysis, 
the approach in which the decay width of the mesons 
may be needed. 
Such calculations will require two-point 
correlation functions of mesons in the Minkowski 
kinematics. 
This is beyond the scope of the present approach. 

As these results have clarified the 
realization of chiral symmetry in the 
scalar mesons, 
it will now be interesting to confirm 
the results experimentally and 
also to study behavior of the 
scalar mesons at finite temperature and/or density, 
where we expect that chiral symmetry is going toward 
restoration.

\newpage

{\Large \bf Acknowledgments }

\vspace{1cm}

I would like to express my deep gratitude for my 
supervisor prof. M. Oka who has 
shared his deep insight in physics with me. 
I have benefited very much from our many 
discussions and his lectures. 
He has also patiently answered my questions. 

I also thank Dr. M. Takizawa for useful discussions 
and comments about the NJL model,
 K. Naito for the discussions  about the ILA approach 
and M. Ishida for useful comment 
 about the scalar nonet mesons.


\appendix

\chapter{BS Equation}

Here we show the explicit form of the BS equation. 
At first, we define the regularized amputated 
BS amplitude $\hat{\chi}^{R}(q;P)$ by 
\begin{eqnarray}
\label{Eq;hatnashi}
\hat{\chi}^{R}(q;P) = 
{S_F^R}^{-1} (q+\frac{P}{2})
\chi^{R}(q;P)
{S_F^R}^{-1} (q-\frac{P}{2})
\end{eqnarray}
By introducing the $\hat{\phi}_A$ which denotes the set of the 
amplitudes of the $\hat{\chi}^{R}(q;P)$ 
, the BS equation can be written as 
\begin{eqnarray}
\label{Eq;hatari}
\hat{\phi}(q;P)  =
\int_k K_{AB} (q,k;P)\phi_B(k;P). 
\end{eqnarray}
We solve the BS equation by dividing to 
these two part. 

The explicit form of Eq. ($\ref{Eq;hatnashi}$) 
is given by 
\begin{eqnarray}
&&\phi_S=\frac{1}{\Delta}
\left[
\begin{array}{l}
\quad \left\{-{q_E}^2+\frac{{P_E}^2}{4}-B_+ B_- \right\}\hat{\phi}_S \\
+\left\{-{q_E}^2(B_+ - B_-) +\frac{{P_E}{q_E}}{2}(B_+ + B_-) \right\}\hat{\phi}_P \\
+\left\{\frac{{P_E}^2}{2}(B_+ + B_-) -{P_E}{q_E}(B_+ - B_-) \right\}\hat{\phi}_Q \\
+\left\{{q_E}^2{P_E}^2-({P_E}{q_E})^2 \right\}\hat{\phi}_T \\
\end{array}
\right]
\\
&&\phi_P=\frac{1}{\Delta}
\left[
\begin{array}{l}
\quad \left\{B_+ -B_- \right\}\hat{\phi}_S \\
+\left\{-{q_E}^2-\frac{{P_E}^2}{4}-B_+ B_- \right\}\hat{\phi}_P \\
+\left\{-2{P_E}{q_E} \right\}\hat{\phi}_Q \\
+\left\{({P_E}{q_E})(B_+ + B_-) -\frac{{P_E}^2}{2}(B_+ - B_-) \right\}\hat{\phi}_T \\
\end{array}
\right]
\\
&&\phi_Q=\frac{1}{\Delta}
\left[
\begin{array}{l}
\quad \left\{-\frac{1}{2}(B_+ +B_-) \right\}\hat{\phi}_S \\
+\left\{\frac{1}{2}{P_E}{q_E}  \right\}\hat{\phi}_P \\
+\left\{\left({q_E}^2+\frac{{P_E}^2}{4}\right)-B_+ B_-\right\}\hat{\phi}_Q \\
+\left\{-{q_E}^2(B_+ + B_-) +\frac{{P_E}{q_E}}{2}(B_+ - B_-) \right\}\hat{\phi}_T \\
\end{array}
\right]
\\
&&\phi_T=\frac{1}{\Delta}
\left[
\begin{array}{l}
\quad \hat{\phi}_S \\
+\left\{\frac{1}{2}(B_+ - B_-)  \right\}\hat{\phi}_P \\
+\left\{B_+ +B_-\right\}\hat{\phi}_Q \\
+\left\{{q_E}^2-\frac{{P_E}^2}{4}-B_+  B_- \right\} \hat{\phi}_T \\
\end{array}
\right]
\end{eqnarray}
where 
\begin{eqnarray}
&& q_{E+} = q+\frac{P}{2}  \qquad  q_{E-} = q-\frac{P}{2}\\
&& B_+ = B (q_{E+}^2) \qquad  B_- = -B(q_{E-}^2)\\
&&\Delta=({q_E}_+^2+B_+^2)({q_E}_-^2+B_-^2). 
\end{eqnarray}

The explicit form of the $\L_{\rm GE}$ part of the
interaction kernel $K$
in the  Eq. ($\ref{Eq;hatnashi}$) 
is given by 
\begin{eqnarray}
&&K^{(E)}_{SS}=C_F\frac{\bar{g}^2}{f^2}\frac{\lambda^{\alpha}}{2}
         \left[
           \frac{-3}{(q_{E}-k_E)^2}
         \right]
          \\
&&K^{(E)}_{PP}=C_F\frac{\bar{g}^2}{f^2}\frac{\lambda^{\alpha}}{2}
         \left[
           \frac{1}{(q_{E}-k_E)^2}\frac{k_q}{A}+\frac{2}{(q_{E}-k_E)^4}(q_{E}-k_E)\cdot k_E
           \left(1-\frac{k_q}{A} \right)
         \right]
         \\
&&K^{(E)}_{QP}=C_F\frac{\bar{g}^2}{f^2}\frac{\lambda^{\alpha}}{2}
         \left[
           \frac{1}{(q_{E}-k_E)^2}\frac{k_P}{A}-\frac{2}{(q_{E}-k_E)^4}(q_{E}-k_E)\cdot k_E
           \left(\frac{k_P}{A} \right)
         \right]
         \\
&&K^{(E)}_{QQ}=C_F\frac{\bar{g}^2}{f^2}\frac{\lambda^{\alpha}}{2}
         \left[
           \frac{1}{(q_{E}-k_E)^2}-\frac{2}{(q_{E}-k_E)^4}(q_{E}-k_E)\cdot P_E
           \left(\frac{k_P}{A} \right)
         \right]
         \\
&&K^{(E)}_{PQ}=C_F\frac{\bar{g}^2}{f^2}\frac{\lambda^{\alpha}}{2}
         \left[
           \frac{2}{(q_{E}-k_E)^4}(q_{E}-k_E)\cdot P_E
           \left(1-\frac{k_q}{A} \right)
         \right]
         \\
&&K^{(E)}_{TT}=C_F\frac{\bar{g}^2}{f^2}\frac{\lambda^{\alpha}}{2}
         \left[
           \frac{ (k_E^2-q_{E}^2)\left(-\frac{k_q}{A} \right)
                 +2(q_{E}-k_E)\cdot P_E \frac{k_P}{A}
                 -2(q_{E}-k_E)\cdot k_E
                }
                {(q_{E}-k_E)^4}
         \right] 
\end{eqnarray}
where
\begin{eqnarray}
&&  A = q_E^2 P_E^2 - (P_E \cdot q_E)^2 \\
&&k_q = P_E^2 (q_E \cdot k_E) - (P_E \cdot q_E)(P_E \cdot k_E) \\
&&k_P = q_E^2 (P_E \cdot k_E) - (q_E \cdot P_E)(q_E \cdot k_E).
\end{eqnarray}

The $\L_{\rm KMT}$ part
is given by 
\begin{eqnarray}
&&
K^{(E)}_{SS} = -24G_D
\epsilon^{ghf'}\epsilon^{fh'g'}
\tr[{S_F}(p)]_{h'h}
\left[
   \frac{\lambda^q}{2}\phi_S^q(q;P) +\frac{\lambda^s}{2}\phi_S^s(q;P)
\right]_{g'h}.
\end{eqnarray}
%

\chapter{Normalization Condition of the BS Amplitude}
Here we show the normalization condition of the BS amplitude 
explicitly. 

To reduce the expressions we use abbreviation 
\begin{eqnarray}
&&B_+ = B\left(\left(q_E+P_E/2\right)^2\right)\\
&&B_- = B\left(\left(q_E-P_E/2\right)^2\right)\\
&&\rho_{+E} = (q_E+P_E/2)^2\\
&&\rho_{-E} = (q_E-P_E/2)^2  \\
&& \Phi = \chi^R(q;P) \\
&& \tilde{\Phi} = \bar{\chi}^R(q;P). 
\end{eqnarray}

The normalization condition of the BS amplitude is 
given by 

\begin{eqnarray}
\label{BSnormalize}
\frac{N_C}{2}
\int_{q_E}&&\frac{1}{f({q_E}_+^2)f({q_E}_-^2)}
\nonumber\\
\times&&
\left\{
 \frac{1}{2}\tr[\tilde{\Phi}\Psla_E\Phi\qsla_E]
-\frac{1}{4}\tr[\tilde{\Phi}\Psla_E\Phi\Psla_E]
-\frac{1}{2}B_- \tr[\tilde{\Phi}\Psla_E\Phi] 
-\left(P_Eq_E+\frac{P_E^2}{2}\right)B'_+\tr[\tilde{\Phi}\Phi\qsla_E]
\right.
\nonumber \\
&& \qquad
+\frac{1}{2}\left( P_Eq_E+\frac{P_E^2}{2} \right)B'_+
\tr[\tilde{\Phi}\Phi\Psla_E]
+\left( P_Eq_E+\frac{P_E^2}{2} \right)B'_+B_- \tr[\tilde{\Phi}\Phi]
\nonumber \\
&&\quad
- \frac{1}{2}\tr[\tilde{\Phi}\qsla_E\Phi\Psla_E]
- \frac{1}{4}\tr[\tilde{\Phi}\Psla_E\Phi\Psla_E]
+ \frac{1}{2}B_+\tr[\tilde{\Phi}\Phi \Psla_E]
-\left(-P_Eq_E+\frac{P_E^2}{2}\right)B'_- \tr[\tilde{\Phi}\qsla_E\Phi]
\nonumber \\
&& \qquad
\left.
-\frac{1}{2}\left(-P_Eq_E+\frac{P_E^2}{2} \right)B'_-
\tr[\tilde{\Phi}\Psla_E\Phi]
+\left(-P_Eq_E+\frac{P_E^2}{2} \right)B_+B'_- \tr[\tilde{\Phi}\Phi]
\right\}
\nonumber \\
&&=2P_E^2
\end{eqnarray}
where
\begin{eqnarray}
\tr[\tilde{\Phi}\Psla_E\Phi\qsla_E]=
&&
4\left(
-\phi_S(-q;P)\phi_S(q;P) P_Eq_E 
-\phi_S(-q;P)\phi_T(q;P) ((P_Eq_E)^2-P_E^2q_E^2)
\right. \nonumber \\
&&
\quad
-\phi_T(-q;P)\phi_S(q;P) ((P_Eq_E)^2-P_E^2q_E^2)
+\phi_P(-q;P)\phi_P(q;P) q_E^2P_Eq_E
\nonumber \\
&&
\quad
+\phi_P(-q;P)\phi_Q(q;P) (P_Eq_E)^2
-\phi_Q(-q;P)\phi_P(q;P) q_E^2P_E^2
\nonumber \\
&&
\quad
\left.
-\phi_Q(-q;P)\phi_Q(q;P) P_E^2P_Eq_E
\right)
\\
\tr[\tilde{\Phi}\qsla_E\Phi\Psla_E]=
&&
4\left(
-\phi_S(-q;P)\phi_S(q;P) P_Eq_E 
+\phi_S(-q;P)\phi_T(q;P) ((P_Eq_E)^2-P_E^2q_E^2)
\right. \nonumber \\
&&
\quad
+\phi_T(-q;P)\phi_S(q;P) ((P_Eq_E)^2-P_E^2q_E^2)
+\phi_P(-q;P)\phi_P(q;P) q_E^2P_Eq_E
\nonumber \\
&&
\quad
+\phi_P(-q;P)\phi_Q(q;P) q_E^2P_E^2
-\phi_Q(-q;P)\phi_P(q;P) (P_Eq_E)^2
\nonumber \\
&&
\quad
\left.
-\phi_Q(-q;P)\phi_Q(q;P) P_E^2P_Eq_E
\right)
\\
\tr[\tilde{\Phi}\Psla_E\Phi\Psla_E]=
&&
4\left(
-\phi_S(-q;P)\phi_S(q;P) P_E^2
-\phi_T(-q;P)\phi_T(q;P) P_E^2\{(P_Eq_E)^2-P_E^2q_E^2\}
\right. \nonumber \\
&&
\quad
+\phi_P(-q;P)\phi_P(q;P) (P_Eq_E)^2
+\phi_P(-q;P)\phi_Q(q;P) P_E^2P_Eq_E
\nonumber \\
&&
\quad
\left.
-\phi_Q(-q;P)\phi_P(q;P) P_E^2P_Eq_E
-\phi_Q(-q;P)\phi_Q(q;P) P_E^4
\right)
\\
\tr[\tilde{\Phi}\Psla_E\Phi]=
&&
-4\left(
-\phi_S(-q;P)\phi_P(q;P) P_Eq_E
-\phi_P(-q;P)\phi_S(q;P) P_Eq_E
\right. \nonumber \\
&&
\quad
-\phi_S(-q;P)\phi_Q(q;P) P_E^2
+\phi_Q(-q;P)\phi_S(q;P) P_E^2
\nonumber\\
&&
\quad
-\phi_T(-q;P)\phi_P(q;P)  \{(P_Eq_E)^2-P_E^2q_E^2\}
\nonumber\\
&&
\quad
\left.
-\phi_P(-q;P)\phi_T(q;P)  \{(P_Eq_E)^2-P_E^2q_E^2\}
\right)
\\
\tr[\tilde{\Phi}\Phi\Psla_E]=
&&
-4\left(
-\phi_S(-q;P)\phi_P(q;P) P_Eq_E
-\phi_P(-q;P)\phi_S(q;P) P_Eq_E
\right. \nonumber \\
&&
\quad
-\phi_S(-q;P)\phi_Q(q;P) P_E^2
+\phi_Q(-q;P)\phi_S(q;P) P_E^2
\nonumber\\
&&
\quad
+\phi_T(-q;P)\phi_P(q;P)  \{(P_Eq_E)^2-P_E^2q_E^2\}
\nonumber\\
&&
\quad
\left.
+\phi_P(-q;P)\phi_T(q;P)  \{(P_Eq_E)^2-P_E^2q_E^2\}
\right)
\\
\tr[\tilde{\Phi}\qsla_E\Phi]=
&&
-4\left(
-\phi_S(-q;P)\phi_P(q;P) q_E^2
-\phi_P(-q;P)\phi_S(q;P) q_E^2
\right. \nonumber \\
&&
\quad
-\phi_S(-q;P)\phi_Q(q;P) P_Eq_E
+\phi_Q(-q;P)\phi_S(q;P) P_Eq_E
\nonumber\\
&&
\quad
+\phi_T(-q;P)\phi_Q(q;P)  \{(P_Eq_E)^2-P_E^2q_E^2\}
\nonumber\\
&&
\quad
\left.
-\phi_Q(-q;P)\phi_T(q;P)  \{(P_Eq_E)^2-P_E^2q_E^2\}
\right)
\\
\tr[\tilde{\Phi}\Phi\qsla_E]=
&&
-4\left(
-\phi_S(-q;P)\phi_P(q;P) q_E^2
-\phi_P(-q;P)\phi_S(q;P) q_E^2
\right. \nonumber \\
&&
\quad
-\phi_S(-q;P)\phi_Q(q;P) P_Eq_E
+\phi_Q(-q;P)\phi_S(q;P) P_Eq_E
\nonumber \\
&&
\quad
-\phi_T(-q;P)\phi_Q(q;P)  \{(P_Eq_E)^2-P_E^2q_E^2\}
\nonumber \\
&&
\quad
\left.
+\phi_Q(-q;P)\phi_T(q;P)  \{(P_Eq_E)^2-P_E^2q_E^2\}
\right)
\\
\tr[\tilde{\Phi}\Phi]=
&&
4\left(
 \phi_S(-q;P)\phi_S(q;P) 
-\phi_P(-q;P)\phi_P(q;P) q_E^2
-\phi_P(-q;P)\phi_Q(q;P) P_Eq_E
\right.
\nonumber \\
&&\quad
+\phi_Q(-q;P)\phi_P(q;P) P_Eq_E
+\phi_Q(-q;P)\phi_Q(q;P) P_E^2
\nonumber \\
&&\quad
\left.
-\phi_T(-q;P)\phi_T(q;P)  \{(P_Eq_E)^2-P_E^2q_E^2\}
\right).
\nonumber \\
\end{eqnarray}

\end{document}